\documentclass[aps,%
twocolumn,%
floatfix,%
pra,%
amsfonts,%
groupedaddress,%
superscriptaddress]{revtex4-2}%
\usepackage[utf8]{inputenc}
\usepackage[T1]{fontenc}
\usepackage{mathrsfs}
\usepackage{mathtools}
\usepackage{amsmath}
\usepackage{amsfonts}
\usepackage{amssymb}
\usepackage[titletoc,title]{appendix}
\usepackage{algorithm,algorithmic}
\makeatletter
\def\ftype@algorithm{8}
\def\endalgorithm{\float@end}
\makeatother

\usepackage{bbm}
\usepackage{bm}
\usepackage{amsthm}
\usepackage{makecell}
\usepackage{array}
\usepackage[dvipsnames,table,xcdraw]{xcolor}
\definecolor{beamer@blendedblue}{rgb}{0.2,0.2,0.7}
\usepackage[osf,sc]{mathpazo}
\usepackage{caption}
\usepackage{subfig}
\usepackage{overpic}
\usepackage{graphicx}
\usepackage{tikz}
\usetikzlibrary{arrows.meta,calc}
\usepackage{booktabs}
\usepackage{multirow}
\newcolumntype{C}[1]{>{\centering\arraybackslash}p{#1}}
\usepackage[colorlinks=true,%
linkcolor=beamer@blendedblue,%
bookmarks=true,%
breaklinks=true,%
filecolor=beamer@blendedblue,%
anchorcolor=yellow,%
citecolor=beamer@blendedblue,%
urlcolor=beamer@blendedblue,%
pdfauthor={Kun Wang},%
pdftitle={Cross-Platform Verification of Continuous-Variable Quantum Systems},%
pdfsubject={Cross-Platform Verification of Continuous-Variable Quantum Systems},%
CJKbookmarks=true]{hyperref}
\usepackage{pifont}

\newtheorem{definition}{Definition}
\newtheorem{proposition}[definition]{Proposition}

\newtheorem{theorem}[definition]{Theorem}
\newtheorem{corollary}[definition]{Corollary}

\mathchardef\ordinarycolon\mathcode`\:
\mathcode`\:=\string"8000
\def\vcentcolon{\mathrel{\mathop\ordinarycolon}}
\begingroup \catcode`\:=\active
  \lowercase{\endgroup
  \let :\vcentcolon
  }

\DeclareFontFamily{U}{mathx}{\hyphenchar\font45}
\DeclareFontShape{U}{mathx}{m}{n}{<-> mathx10}{}
\DeclareSymbolFont{mathx}{U}{mathx}{m}{n}
\DeclareMathAccent{\widebar}{0}{mathx}{"73}

\newcommand{\wh}[1]{\widehat{#1}}
\newcommand{\summary}[1]{\medskip\textbf{#1}}

\newcommand{\ket}[1]{\vert{#1}\rangle}
\newcommand{\bra}[1]{\langle{#1}\vert}
\newcommand{\ketbra}[1]{\vert{#1}\rangle\!\langle{#1}\vert}

\newcommand{\density}[1]{\mathscr{D}(#1)}

\DeclareMathOperator{\tr}{Tr}
\newcommand{\1}{\mathbbm{1}}

\def\ve{\varepsilon}
\newcommand{\norm}[2]{\ensuremath{\left\lVert#1\right\rVert_{#2}}}%

\newcommand{\Var}{\mathbb{V}}

\makeatletter
\newsavebox{\@brx}
\newcommand{\llangle}[1][]{\savebox{\@brx}{\(\m@th{#1\langle}\)}%
  \mathopen{\copy\@brx\kern-0.5\wd\@brx\usebox{\@brx}}}
\newcommand{\rrangle}[1][]{\savebox{\@brx}{\(\m@th{#1\rangle}\)}%
  \mathclose{\copy\@brx\kern-0.5\wd\@brx\usebox{\@brx}}}
\makeatother

\newcommand*{\cC}{\mathcal{C}}

\newcommand*{\cH}{\mathcal{H}}

\newcommand*{\cO}{\mathcal{O}}

\newcommand*{\cT}{\mathcal{T}}

\newcommand{\bE}{\mathbb{E}}
\newcommand{\bC}{\mathbb{C}}

\newcommand{\bR}{\mathbb{R}}
\newcommand{\bP}{\mathbb{P}}

\newcommand{\bN}{\mathbb{N}}

\newcommand{\Piop}{\Pi}

\begin{document}

\newcommand{\thetitle}{{Cross-Platform Verification of Continuous-Variable Quantum Systems}}
\title{\thetitle}

\author{Kun Wang}
\email{nju.wangkun@gmail.com}
\affiliation{College of Computer Science and Technology, 
National University of Defense Technology, 
Changsha 410073, China}%

\date{\today}


\begin{abstract}
Cross-platform verification compares output states of independently
operated bosonic and quantum optical systems when classical predictions
and joint measurements are unavailable. Continuous-variable distributed
inner product estimation (CV-DIPE) provides a local measurement approach,
but infinite-dimensional state spaces require rigorous control of truncation error.
We develop three CV-DIPE protocols for the natural class of
energy-constrained continuous-variable quantum states. One uses
coordinated displaced parity, another coordinated Haar measurements, and
the third independent homodyne shadows. For arbitrary multimode inputs in this class,
including entangled and non-Gaussian states, 
we establish truncation error bounds and worst-case sample complexity guarantees for the exact inner product.
The leading terms of these bounds favor coordinated Haar measurements,
whose global implementation is experimentally challenging. 
These protocols provide certified strategies for cross-platform
verification of continuous-variable systems using only local measurement data.
\end{abstract}

\maketitle

\section{Introduction}

Many benchmarks compare a quantum processor's output with a classical prediction, but computing that prediction becomes difficult for experiments beyond efficient classical simulation~\cite{eisert2020quantuma}. 
Cross-platform verification directly compares states prepared by
independent devices, quantifying their agreement without knowing either
state in advance.
Such comparisons have been studied and demonstrated on trapped ion and superconducting processors and across distinct platforms~\cite{elben2020cross,zhu2022cross,dalton2025resourceefficient,wu2025state,zheng2024crossplatform}.
Distributed inner product estimation (DIPE) formalizes this task~\cite{carrasco2021theoretical,elben2022randomized}. Two laboratories estimate the inner product of their unknown states using only local measurements and classical communication.
Continuous-variable (CV) quantum computing is a major approach
to quantum computation. Advances in bosonic devices make cross-platform
verification increasingly relevant. Integrated photonic systems have demonstrated multipartite entanglement~\cite{jia2025microcomb} and the generation, manipulation, and measurement of cluster states~\cite{jia2026monolithic}, while optical experiments have prepared cat states in propagating light~\cite{konno2024logical} and Gottesman--Kitaev--Preskill (GKP) states~\cite{larsen2025gkp}. Modular photonic processors~\cite{aghaee2025modular} and nonlinear optical operations~\cite{yu2026nonlinearity} further broaden the approaches to bosonic quantum computing.
CV-DIPE thus provides a route to cross-platform verification of CV systems.
It quantifies agreement between states prepared by independently
operated processors when classical predictions are infeasible and the
states cannot be brought together.

In finite dimensions, the copy complexity of DIPE is sharply characterized even with adaptive local measurements on multiple copies and unrestricted classical communication~\cite{anshu2022distributed}; structured random circuits offer further measurement designs~\cite{zheng2025distributed,huang2026worst}. These results do not automatically extend to bosonic inputs, whose phase space is unbounded. Joint bosonic interference permits inner product estimation with a sample count independent of dimension and has been implemented for integrated photonic qudit states, but requires both inputs at a common measurement device~\cite{volkoff2022,zhan2025experimental}. Wu \emph{et al.} used a convolutional neural network trained on related states to compare unknown CV states from limited, noisy measurement data~\cite{wu2023similarity}. Li \emph{et al.} analyzed a distributed method based on characteristic functions for self-reflective CV states, subject to an additional condition on their phase space tails~\cite{li2026machine}. Fidelity estimation using displaced parity measurements with a known target state~\cite{fawzi2026optimal} and CV shadow methods~\cite{gandhari2024precision,iosue2024continuous,becker2024classical,thomas2025shedding} address related questions under different input or measurement models.
For unknown, separated bosonic states, comparing distinct
local measurement strategies requires rigorous control of truncation
error and sample complexity under an explicit physical constraint.

\textbf{Contributions.} We focus on arbitrary, possibly entangled and non-Gaussian multimode states under a known energy constraint~\cite{mele2025}.
For this input class, we develop three CV-DIPE protocols based on coordinated displaced parity, coordinated Haar measurements, and independent homodyne shadows. We bound the bias from phase space or Fock truncation and prove worst-case sample complexity guarantees for estimating the exact inner product. Comparing these guarantees and the required measurements yields qualitative observations that may help select a suitable CV-DIPE protocol.

\textbf{Organization.} Secs.~\ref{sec:dipe-cv}--\ref{sec:cv-dipe-shadows} develop the three CV-DIPE protocols based on coordinated displaced parity, coordinated Haar measurements, and independent homodyne shadows, respectively. Sec.~\ref{sec:comparison} compares their guarantees and experimental requirements. Sec.~\ref{sec:conclusion} concludes. The Appendix provides proofs.

\section{CV-DIPE via Coordinated Displaced Parity}
\label{sec:dipe-cv}

This section develops the coordinated displaced parity protocol on the energy-constrained class. 
We define the task, certify a finite ball in phase space and its truncation bias from the energy promise, 
then analyze the estimator and its worst-case sample complexity.

\subsection{The CV-DIPE task}

CV-DIPE asks two separated laboratories to estimate the exact inner product of unknown states 
on the $m$-mode bosonic Fock space $\cH_m$ using only local quantum operations and classical communication. 
Alice holds copies of $\rho$ and Bob holds copies of $\sigma$; 
they may share randomness and exchange classical messages, 
but cannot exchange quantum systems or use preshared entanglement. 
Let $\density{\cH_m}$ denote the set of density operators on $\cH_m$. 
For a state class $\cC\subset\density{\cH_m}$ and $\ve,\delta\in(0,1)$, 
the task is to construct an estimator $\widehat s$ such that
\begin{align}
\sup_{\rho,\sigma\in\cC}
\bP\!\left(\left|\widehat s-\tr(\rho\sigma)\right|>\ve\right)
\leq\delta.
\label{eq:DIPE}
\end{align}

When both state preparations can be brought to one apparatus, an ideal joint SWAP measurement provides a benchmark for bosonic inner product estimation~\cite{daley2012,volkoff2022,descamps2026}. Interference followed by photon counting or parity readout can implement the exchange measurement. Repeated SWAP measurements meet the accuracy and confidence requirements of Eq.~\eqref{eq:DIPE} with $O(1/\ve^2\ln(1/\delta))$ copies, independent of the Hilbert space dimension; the dependence on inverse accuracy is optimal in the worst case at constant confidence even for general joint measurements~\cite{anshu2022distributed}. This benchmark requires the two states to meet and therefore does not apply when they remain at separate laboratories.

Two measures for comparing mixed states that depend on the inner product are the max fidelity defined by Liang \emph{et al.}~\cite{liang2019} and later used for verification across platforms by Elben \emph{et al.}~\cite{elben2020cross}, and the squared Hilbert--Schmidt distance~\cite{anshu2022distributed}:
\begin{align}
F_{\max}(\rho,\sigma)
&=\frac{\tr(\rho\sigma)}
{\max\{\tr(\rho^2),\tr(\sigma^2)\}},\\
D_{\mathrm{HS}}^2(\rho,\sigma)
&=\tr(\rho^2)+\tr(\sigma^2)-2\tr(\rho\sigma).
\end{align}
A CV-DIPE protocol that estimates the inner product can also estimate each purity by taking its two inputs to be independent copies of the same state. Both quantities can then be estimated by substituting the inner product and purity estimates into the formulas above, without reconstructing either state.

The Wigner function provides a real phase space representation of a bosonic state~\cite{weedbrook2012gaussian}. 
Let $\bC$ denote the complex numbers.
For a state $\tau$ on the $m$-mode Fock space $\cH_m$, write $W_\tau(\boldsymbol{\alpha})$ 
for the Wigner function of $\tau$, with $\boldsymbol{\alpha}=(\alpha_1,\dots,\alpha_m)\in\bC^m$ 
specifying a phase space point. 
The Wigner inner product identity~\cite{fawzi2026optimal} reads
\begin{align}
\tr(\rho\sigma)
=\pi^m\int_{\bC^m}W_\rho(\boldsymbol{\alpha})W_\sigma(\boldsymbol{\alpha})\,d\boldsymbol{\alpha}.
\label{eq:overlap-wigner}
\end{align}
This identity suggests a Monte Carlo estimate obtained by sampling common phase space points $\boldsymbol{\alpha}$ and averaging appropriately weighted estimates of the product of the two Wigner functions~\cite{robert2004montecarlointegration}. To implement this proposal with uniform draws and local measurements, two obstacles must be resolved:
\begin{enumerate}
  \item the space $\bC^m$ has infinite volume and therefore admits no normalized uniform distribution, and 
  \item the identity provides no local measurement procedure for either Wigner value. 
\end{enumerate}

We address these obstacles in turn by truncating phase space and using displaced parity measurements.
We denote the tolerated truncation bias by $\ve_{\rm tail}$ and the tolerated
statistical deviation of the estimator from the truncated inner product by
$\ve_{\rm stat}$. These two error budgets satisfy
$\ve_{\rm tail}+\ve_{\rm stat}=\ve$.

\subsection{Certified truncation in phase space}
\label{subsec:truncated-tail}
\label{subsec:energy-region}

We address the first obstacle by restricting the integral to a measurable
region $\Omega\subset\bC^m$ with positive finite volume,
\begin{align}
0<|\Omega|:=\int_\Omega d\boldsymbol{\alpha}<\infty,
\end{align}
where $d\boldsymbol{\alpha}=d^{2m}\boldsymbol{\alpha}$ is Lebesgue measure. The density $1/|\Omega|$ on $\Omega$ defines the uniform distribution~\cite[Example~1.75]{klenke2020probability}. Multiplying the Wigner product under this distribution by $\pi^m|\Omega|$ gives the \emph{truncated inner product}
\begin{align}
\cT_\Omega(\rho,\sigma):=\pi^m\int_\Omega W_\rho(\boldsymbol{\alpha})W_\sigma(\boldsymbol{\alpha})\,d\boldsymbol{\alpha}.
\label{eq:truncated-inner-product}
\end{align}

The \emph{truncation bias} is the error from omitting the region outside $\Omega$, namely $\left|\tr(\rho\sigma)-\cT_\Omega(\rho,\sigma)\right|$. Finite volume permits uniform sampling but does not ensure a small bias. The following proposition makes this limitation precise for unrestricted CV states; its proof is in Appx.~\ref{appx:proof-of-proposition-impossible}.
\begin{proposition}
\label{prop:impossible}
Let $\Omega\subset\bC^m$ be measurable with $|\Omega|<\infty$. Then
\begin{align}\label{eq:impossible}
\sup_{\rho\in\density{\cH_m}}\left|\tr(\rho^2)-\cT_\Omega(\rho,\rho)\right|=1.
\end{align}
\end{proposition}

A sequence of increasingly displaced coherent states approaches the worst case in Prop.~\ref{prop:impossible}. For this sequence, the truncated inner product of each state with itself tends to zero while its purity remains one, even if $\Omega$ is unbounded but has finite volume.
Therefore, a uniform bias bound for a fixed region 
requires a promise on the state class that controls the omitted Wigner contribution.

\summary{Energy-constrained states.}
We adopt the energy constraint of Ref.~\cite{mele2025}. It reflects the
finite energy of experimentally prepared optical states and can often be
supported by prior knowledge and characterization of the source. A mean
energy bound also controls the unobserved phase space and Fock tails,
allowing finite truncations with certified bias. For each
mode $j$, let $a_j$ and $a_j^\dagger$ denote its annihilation and
creation operators, respectively. Let $E\geq1/2$ be the energy
constraint, and define
\begin{align}
\hat\nu_{\rm tot}&:=\sum_{j=1}^m a_j^\dagger a_j,\qquad
\hat E_m:=\hat\nu_{\rm tot}+\frac m2\1,\\
\cC_{m,E}&:=\left\{\tau\in\density{\cH_m}:
\tr(\hat E_m\tau)\leq mE\right\}.
\label{eq:energy-bounded-state}
\end{align}
The vacuum has energy $m/2$, the minimum possible value of
$\tr(\hat E_m\tau)$. Thus $E\geq1/2$ is necessary for the class to
be nonempty; at $E=1/2$, it contains only the multimode vacuum.
Equivalently, the promise is
$\tr(\hat\nu_{\rm tot}\tau)\leq m(E-1/2)$.
It permits arbitrary correlations and energy distributions among
the modes.

For $R>0$, define the centered ball
\begin{align}
B_R:=\left\{\boldsymbol\alpha\in\bC^m:
\sum_{j=1}^m|\alpha_j|^2\leq R^2\right\}.
\label{eq:energy-region}
\end{align}
Its volume is $|B_R|=\pi^mR^{2m}/m!$.
To quantify the omitted Wigner contribution, let
$B_R^c:=\bC^m\setminus B_R$ and let $\mathbbm{1}_{B_R^c}$ denote its
indicator function. Define the \emph{Wigner tail norm} of a state
$\tau$ outside $B_R$ by
\begin{align}\label{eq:energy-Wigner-tail-norm}
\norm{W_\tau\mathbbm{1}_{B_R^c}}{2}
\coloneqq\left(\int_{B_R^c}|W_\tau(\boldsymbol{\alpha})|^2\,d\boldsymbol{\alpha}\right)^{1/2}.
\end{align}
This norm measures the omitted Wigner function without cancellation
between positive and negative values.
The bound on the second moment of the Wigner function yields
\begin{align}
\sup_{\tau\in\cC_{m,E}}
\norm{W_\tau\mathbbm{1}_{B_R^c}}{2}
\leq\eta_{B_R}:=\sqrt{\frac{mE}{\pi^mR^2}}.
\label{eq:energy-wigner-tail}
\end{align}
This tail bound controls the truncation bias uniformly.
\begin{proposition}
\label{prop:energy-truncation-bias}
For any $R>0$,
\begin{align}
\sup_{\rho,\sigma\in\cC_{m,E}}
\left|\tr(\rho\sigma)-\cT_{B_R}(\rho,\sigma)\right|
\leq\pi^m\eta_{B_R}^2 = mE/R^2.
\label{eq:energy-overlap-bias}
\end{align}
\end{proposition}
Eq.~\eqref{eq:energy-wigner-tail} and
Prop.~\ref{prop:energy-truncation-bias} are proved in
Appx.~\ref{appx:energy-tail-bias}.
To meet the truncation error budget, choose $R$ such that
$mE/R^2\leq\ve_{\rm tail}$.

\summary{Estimating the energy constraint.}
Separate characterization of Alice's and Bob's sources can provide
certified upper bounds $M_\rho$ and $M_\sigma$ on their mean total photon
numbers. Any common parameter
$E\geq1/2+\max\{M_\rho,M_\sigma\}/m$ then certifies
$\rho,\sigma\in\cC_{m,E}$. Calibrated photon counting or homodyne
measurements~\cite{webb2006homodyne} can estimate the required means,
with loss, mode matching, and calibration uncertainty accounted for.
However, finite data alone cannot upper bound an unrestricted mean, since rare
events with high photon occupation may be missed; a justified source model or tail
constraint is also needed.

\begin{figure*}[t]
\centering
\definecolor{protocolInk}{HTML}{5F6368}
\definecolor{protocolBlue}{HTML}{4285F4}
\definecolor{protocolRed}{HTML}{DB4437}
\definecolor{protocolYellow}{HTML}{F4B400}
\definecolor{protocolGreen}{HTML}{0F9D58}
\definecolor{protocolBackground}{HTML}{F1F3F4}
\colorlet{protocolPanel}{protocolBlue!4!protocolBackground}
\def\protocolWireLength{0.40cm}
\begin{tikzpicture}[
  x=1cm,y=1cm,
  every node/.style={font=\normalfont\normalsize},
  stage/.style={draw=protocolInk!55!protocolBackground,fill=protocolBackground,rounded corners=2pt,
    minimum width=17.10cm,line width=0.6pt},
  lab/.style={draw=protocolInk!45!protocolBackground,fill=protocolPanel,rounded corners=2pt,
    minimum width=7.95cm,minimum height=1.80cm,line width=0.6pt},
  subbox/.style={minimum width=1.85cm,minimum height=0.95cm,
    rounded corners=1.5pt,line width=0.6pt,align=center},
  outcome/.style={minimum width=0.60cm,minimum height=0.60cm,inner sep=2pt},
  platform/.style={subbox,draw=protocolRed!75!protocolInk,fill=protocolRed!8!protocolPanel},
  controlled/.style={subbox,draw=protocolBlue!80!protocolInk,fill=protocolBlue!11!protocolPanel},
  detector/.style={subbox,draw=protocolGreen!80!protocolInk,fill=protocolGreen!9!protocolPanel},
  wire/.style={-{Latex[length=1.5mm,width=1.0mm]},draw=protocolInk,
    solid,line width=0.65pt},
  quantum/.style={wire},
  classical/.style={wire,dash pattern=on 3pt off 1.8pt}
]
\path[use as bounding box] (-0.04,0.61) rectangle (17.14,8.32);
\node[stage,minimum height=1.15cm]
  (region) at (8.55,7.705) {};
\node[font=\normalfont\normalsize\bfseries] at (8.55,7.93)
  {Stage 1. Region selection};
\node at (8.55,7.46)
  {Select the centered ball $B_R$ with $R^2=2mE/\ve$};

\node[stage,minimum height=5.55cm]
  (estimation) at (8.55,3.425) {};
\node[font=\normalfont\normalsize\bfseries] at (8.55,5.87)
  {Stage 2. Truncated inner product estimation};
\draw[classical] (region.south) --
  node[right,inner sep=3pt] {parity rounds $N$} (estimation.north);

\node[lab] (alice) at (4.325,3.07) {};
\node[lab] (bob) at (12.775,3.07) {};
\node[font=\normalfont\normalsize\bfseries]
  at (7.10,3.66) {Alice};
\node[font=\normalfont\normalsize\bfseries]
  at (10.00,3.66) {Bob};

\node[platform]
  (asrc) at (7.10,2.80) {Bosonic $\rho$};
\node[controlled,anchor=east]
  (adisp) at ($(asrc.west)+(-\protocolWireLength,0)$) {$D(-\boldsymbol{\alpha}_t)$};
\node[detector,anchor=east]
  (adet) at ($(adisp.west)+(-\protocolWireLength,0)$) {PNRD};
\draw[quantum] (asrc.west) -- (adisp.east);
\draw[quantum] (adisp.west) -- (adet.east);
\node[outcome,anchor=east]
  (aout) at ($(adet.west)+(-\protocolWireLength,0)$) {$\bm{k}_t$};
\draw[classical] (adet.west) -- (aout.east);

\node[platform]
  (bsrc) at (10.00,2.80) {Bosonic $\sigma$};
\node[controlled,anchor=west]
  (bdisp) at ($(bsrc.east)+(\protocolWireLength,0)$) {$D(-\boldsymbol{\alpha}_t)$};
\node[detector,anchor=west]
  (bdet) at ($(bdisp.east)+(\protocolWireLength,0)$) {PNRD};
\draw[quantum] (bsrc.east) -- (bdisp.west);
\draw[quantum] (bdisp.east) -- (bdet.west);
\node[outcome,anchor=west]
  (bout) at ($(bdet.east)+(\protocolWireLength,0)$) {$\bm{\ell}_t$};
\draw[classical] (bdet.east) -- (bout.west);

\node[draw=protocolYellow!65!protocolInk,fill=protocolYellow!16!protocolBackground,rounded corners=2pt,
  line width=0.6pt,inner xsep=8pt,inner ysep=5pt,align=center]
  (seed) at (8.55,5.09)
  {Sample $\boldsymbol{\alpha}_t\sim\mathrm{Unif}(B_R)$};
\coordinate (shared-split) at (8.55,4.42);
\draw[classical,-] (seed.south) -- (shared-split);
\draw[classical] (shared-split) -| (adisp.north);
\draw[classical] (shared-split) -| (bdisp.north);

\node[draw=protocolInk!45!protocolBackground,fill=protocolPanel,rounded corners=2pt,
  minimum width=16.40cm,minimum height=0.95cm,line width=0.6pt]
  (post) at (8.55,1.325) {};
\draw[classical] (aout.south) -- (aout.south |- post.north);
\draw[classical] (bout.south) -- (bout.south |- post.north);
\node at (post.center)
  {$\wh{\cT}_{B_R}=(4/\pi)^m|B_R|
    \sum_{t=1}^{N}(-1)^{|\bm{k}_t|}(-1)^{|\bm{\ell}_t|}/N
    \approx\tr(\rho\sigma)$};
\end{tikzpicture}
\caption{\raggedright\textbf{CV-DIPE via coordinated displaced parity.}
Stage~1 selects the ball $B_R$ from the energy bound, with $R^2=2mE/\ve$.
Its volume $|B_R|$ determines the sufficient
number $N$ of parity rounds at statistical tolerance $\ve/2$
and failure probability $\delta$. In Stage~2,
Alice and Bob sample common points uniformly from $B_R$, apply the
corresponding displacements to fresh copies of their states, and measure Fock
occupations with detectors that resolve photon number (PNRDs). The rescaled
average product of their total parities yields $\wh{\cT}_{B_R}$, which
estimates $\tr(\rho\sigma)$ with controlled truncation bias and statistical error.}
\label{fig:cv-dipe-protocol}
\end{figure*}
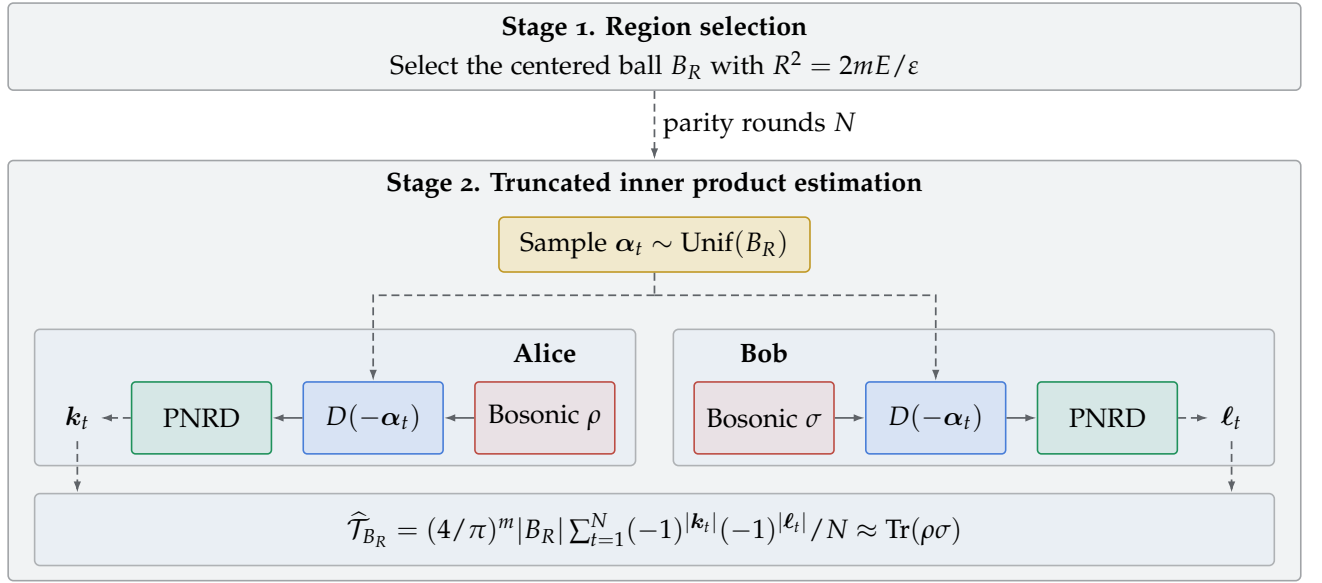

\subsection{Truncated inner product estimation}

We address the second obstacle by using local displaced parity measurements
to estimate the Wigner values needed for $\cT_{B_R}(\rho,\sigma)$.
Using the certified ball $B_R$, we construct
an estimator and bound its statistical deviation by $\ve_{\rm stat}$.

The expectation of displaced parity determines each Wigner value in the truncated integral. Let $\alpha_j^*$ denote the complex conjugate of $\alpha_j$. Define the single-mode displacement, multimode displacement, and total parity operators by
\begin{align}
D(\alpha_j)&\coloneqq\exp\left(\alpha_j a_j^\dagger-\alpha_j^*a_j\right),\\
D(\boldsymbol{\alpha})&\coloneqq\bigotimes_{j=1}^m D(\alpha_j),\\
\Piop&\coloneqq(-1)^{\hat \nu_{\rm tot}}.
\end{align}
The displaced parity representation~\cite{fawzi2026optimal} reads
\begin{align}
W_\tau(\boldsymbol{\alpha})
=\left(\frac{2}{\pi}\right)^m
\tr\!\left(\tau D(\boldsymbol{\alpha})\Piop D(\boldsymbol{\alpha})^\dagger\right).
\label{eq:wigner-parity}
\end{align}
Operationally, one estimates $W_\tau(\boldsymbol{\alpha})$ by displacing copies of $\tau$ by $-\boldsymbol{\alpha}$, measuring their Fock occupations, and averaging the parity of the total occupation. Each parity outcome has mean $(\pi/2)^m W_\tau(\boldsymbol{\alpha})$, so multiplying the average by $(2/\pi)^m$ gives the Wigner estimate.

The truncated inner product $\cT_{B_R}(\rho,\sigma)$
can then be estimated by sampling common points uniformly from $B_R$ and rescaling the average product
of local parity outcomes. Let $\bN_0:=\{0,1,2,\dots\}$ denote the nonnegative integers.
For a Fock occupation vector $\bm{k}=(k_1,\dots,k_m)\in\bN_0^m$, define its total photon number $|\bm{k}|$ and the total parity function $s(\bm{k})$ by
\begin{align}
|\bm{k}|&\coloneqq\sum_{j=1}^m k_j,\qquad
s(\bm{k})\coloneqq(-1)^{|\bm{k}|}.
\end{align}
Let $N$ be the number of \emph{parity rounds}, indexed by $t=1,\dots,N$. Each round consists of sampling a common point $\boldsymbol{\alpha}_t$ uniformly from $B_R$, displacing one fresh copy of each state by $-\boldsymbol{\alpha}_t$, and measuring both copies in the multimode Fock basis. Alice and Bob obtain $\bm{k}_t$ and $\bm{\ell}_t$, respectively, and record $X_t=s(\bm{k}_t)$ and $Y_t=s(\bm{\ell}_t)$. We construct the following estimator
\begin{align}
\wh{\cT}_{B_R}:=\left(\frac{4}{\pi}\right)^m |B_R|\,\frac{1}{N}\sum_{t=1}^N X_tY_t.
\label{eq:estimator-main}
\end{align}

We write $\bE[\cdot]$ for expectation over measurement settings and
outcomes and $\bE[\cdot\mid\boldsymbol{\alpha}_t=\boldsymbol{\alpha}]$
for expectation conditional on the sampled point.
This estimator is unbiased for the truncated inner product. At a fixed sampled point $\boldsymbol{\alpha}$, Eq.~\eqref{eq:wigner-parity} gives mean parity $\left(\pi/2\right)^mW_\rho(\boldsymbol{\alpha})$ for Alice and $\left(\pi/2\right)^mW_\sigma(\boldsymbol{\alpha})$ for Bob. Their outcomes are independent conditioned on that point, so averaging their product over the uniform draw gives
\begin{align}
\bE[\wh{\cT}_{B_R}]
&=\left(\frac{4}{\pi}\right)^m\int_{B_R}
\bE\left[X_tY_t\mid\boldsymbol{\alpha}_t=\boldsymbol{\alpha}\right]d\boldsymbol{\alpha}\\
&=\pi^m\int_{B_R} W_\rho(\boldsymbol{\alpha})W_\sigma(\boldsymbol{\alpha})\,d\boldsymbol{\alpha}
=\cT_{B_R}(\rho,\sigma).
\label{eq:unbiased-truncated}
\end{align}

Furthermore, the parity outcomes $X_t,Y_t\in\{-1,1\}$ are bounded for every input state. Hoeffding's inequality therefore gives the following nonasymptotic sample complexity guarantee for the truncated inner product,
whose proof is given in Appx.~\ref{appx:proof-of-theorem-hoeffding}.
 
\begin{theorem}
\label{thm:hoeffding}
For any $\ve_{\rm stat},\delta\in(0,1)$, if
\begin{align}
N\ge 2\left(\frac{4}{\pi}\right)^{2m}\frac{|B_R|^2}{\ve_{\rm stat}^2}\log\frac{2}{\delta},
\label{eq:M-hoeffding}
\end{align}
then the estimator in Eq.~\eqref{eq:estimator-main} satisfies
\begin{align}
\sup_{\rho,\sigma\in\density{\cH_m}}
\bP\!\left(\left|\wh{\cT}_{B_R}-\cT_{B_R}(\rho,\sigma)\right|\ge \ve_{\rm stat}\right)\le \delta.
\end{align}
\end{theorem}

\subsection{CV-DIPE via coordinated displaced parity}
\label{subsec:parity-protocol-energy}

\begin{algorithm}
\caption{CV-DIPE via coordinated displaced parity}
\label{alg:cv-dipe}
\begin{algorithmic}[1]
\REQUIRE Unknown $m$-mode states $\rho,\sigma\in\cC_{m,E}$;
energy bound $E$; accuracy $\ve\in(0,1)$ and
failure probability $\delta\in(0,1)$.
\ENSURE An estimate of $\tr(\rho\sigma)$ with additive error at most
$\ve$, except with probability $\delta$.
\STATE Set $R^2=2mE/\ve$ and select the ball $B_R$ of Eq.~\eqref{eq:energy-region}.
\STATE Choose $N$ parity rounds by Eq.~\eqref{eq:energy-copy-complexity}.
\FOR{$t=1,\ldots,N$}
\STATE Sample $\boldsymbol\alpha_t\sim\mathrm{Unif}(B_R)$ using
shared randomness.
\STATE Alice applies $D(-\boldsymbol\alpha_t)$ to a fresh copy of
$\rho$ and measures Fock occupations, obtaining $\bm k_t$.
\STATE Bob applies $D(-\boldsymbol\alpha_t)$ to a fresh copy of
$\sigma$ and measures Fock occupations, obtaining $\bm\ell_t$.
\STATE Record $X_t=s(\bm k_t)$ and $Y_t=s(\bm\ell_t)$.
\ENDFOR
\STATE \textbf{Return} $\wh{\cT}_{B_R}$ from
Eq.~\eqref{eq:estimator-main}.
\end{algorithmic}
\end{algorithm}

The certified ball of Eq.~\eqref{eq:energy-region} and the displaced
parity estimator of Eq.~\eqref{eq:estimator-main} together give an
end-to-end CV-DIPE protocol for energy-constrained states.
Alg.~\ref{alg:cv-dipe} presents the protocol, and
Fig.~\ref{fig:cv-dipe-protocol} illustrates its measurement scheme.
We allocate the target error equally between truncation and sampling, with
$\ve_{\rm tail}=\ve_{\rm stat}=\ve/2$.
Choosing $R^2=2mE/\ve$ bounds the truncation bias by $\ve/2$, while
the volume of $B_R$ determines the number of parity rounds needed to
control the statistical error. This yields the following worst-case
sample complexity guarantee for the exact inner product.

\begin{corollary}
\label{cor:energy-end-to-end}
For $\rho,\sigma\in\cC_{m,E}$ and
$\ve,\delta\in(0,1)$, choose $R^2=2mE/\ve$. If the number of copies
per party satisfies
\begin{align}
N\geq\frac{8}{\ve^2}
\left(\frac{8eE}{\ve}\right)^{2m}
\log\frac{2}{\delta},
\label{eq:energy-copy-complexity}
\end{align}
then the output of Alg.~\ref{alg:cv-dipe} obeys
\begin{align}
\sup_{\rho,\sigma\in\cC_{m,E}}
\bP\!\left(
\left|\wh{\cT}_{B_R}-\tr(\rho\sigma)\right|>\ve
\right)\leq\delta.
\end{align}
\end{corollary}
The corollary follows directly from
Prop.~\ref{prop:energy-truncation-bias} and
Thm.~\ref{thm:hoeffding}.
Each parity round consumes one copy per party. For fixed $m$ and $E$,
the sufficient scaling is
$\cO(\ve^{-2m-2}\log(2/\delta))$ copies per party. The guarantee
holds for all, including entangled, states in $\cC_{m,E}$.

\section{CV-DIPE via Coordinated Haar Measurements}
\label{sec:fock-dv-dipe}

The energy constraint permits an approximate reduction of CV-DIPE to
finite-dimensional DIPE with a uniform bound on the truncation error.
For arbitrary, possibly entangled, $m$-mode inputs in $\cC_{m,E}$,
we first certify a Fock cutoff and then apply the coordinated Haar
measurement protocol of Ref.~\cite{anshu2022distributed} (hereafter the
DV-DIPE protocol) to the resulting finite-dimensional states. The
optimality result of Ref.~\cite{anshu2022distributed} concerns unrestricted
finite-dimensional DIPE; the resulting CV bound is an upper bound on
worst-case sample complexity under the energy constraint.

\summary{Certified Fock cutoff.}
Let $\ket{n}$ denote the Fock state of one mode with $n$ photons.
For an integer $L\geq1$, let $d_L:=L^m$ and define
\begin{subequations}\label{eq:fock-energy-cutoff}
\begin{align}
P_L
&:=\bigotimes_{j=1}^{m}\sum_{n=0}^{L-1}\ketbra{n}, \\
q_L
&:=\min\left\{1,\frac{mE}{L}\right\}.
\end{align}  
\end{subequations}
If a Fock occupation outcome lies outside $P_L$, at least one mode
contains $L$ or more photons.
Since $\tr(\hat\nu_{\rm tot}\tau)\leq mE$, Markov's inequality gives
$\tr[(\1-P_L)\tau]\leq q_L$
for every $\tau\in\cC_{m,E}$. The Hilbert--Schmidt bound
proved in Appx.~\ref{appx:fock-cutoff-overlap}
then gives
\begin{align}
\sup_{\rho,\sigma\in\cC_{m,E}}
\left|\tr(\rho\sigma)
-\tr(P_L\rho P_L
P_L\sigma P_L)\right|
\leq 2q_L.
\label{eq:fock-dv-bias}
\end{align}
The certificate is uniform over the promised class, including states
entangled across modes.

\summary{DV-DIPE protocol.}
The projected operators $P_L\rho P_L$ and $P_L\sigma P_L$ are
subnormalized. To apply the finite-dimensional DV-DIPE protocol to
density operators without conditioning on successful truncation, we
adjoin two orthogonal flags $\ket{a}$ and $\ket{b}$ to the range of
$P_L$. With $Q_L=\1-P_L$,
Alice and Bob locally prepare, respectively,
\begin{subequations}\label{eq:fock-dv-flag-states}
\begin{align}
\rho_L^{A}
&:=P_L\rho P_L
+\tr(Q_L\rho)\ketbra{a},\\
\sigma_L^{B}
&:=P_L\sigma P_L
+\tr(Q_L\sigma)\ketbra{b}.
\end{align}
\end{subequations}
These are normalized states on a common space of dimension
$D_L=d_L+2$. Each map is implemented locally
by a nondemolition $\{P_L,Q_L\}$ measurement:
retain the state on the first outcome and replace the second by the
party's flag. Since the flags are orthogonal to each other and to the
Fock sector,
\begin{align}
\tr(\rho_L^{A}\sigma_L^{B})
=\tr(P_L\rho P_L
P_L\sigma P_L).
\label{eq:fock-dv-flag-overlap}
\end{align}

Alice and Bob run the DV-DIPE protocol on $\rho_L^A$ and $\sigma_L^B$
using shared measurements drawn from the Haar distribution in dimension $D_L$. The protocol
returns an estimator $\wh s_{\rm Haar}$ of the inner product in
Eq.~\eqref{eq:fock-dv-flag-overlap}. For statistical error at most
$\ve_{\rm stat}$ and failure probability $\delta$,
Ref.~\cite{anshu2022distributed} gives a sufficient count of
$\cO(\log(2/\delta)\max\{\ve_{\rm stat}^{-2},
\sqrt{D_L}/\ve_{\rm stat}\})$
copies per party. The laboratories exchange only classical data.

The cutoff used below gives $q_L\leq\ve/4$, so
Eq.~\eqref{eq:fock-dv-bias} bounds the bias by $\ve/2$. Taking
$\ve_{\rm stat}=\ve/2$ in the DV-DIPE guarantee bounds the statistical
error by $\ve/2$ with failure probability $\delta$. Since
$d_L\leq(1+4mE/\ve)^m$, the resulting sufficient count is
Eq.~\eqref{eq:fock-dv-explicit-count}.
Alg.~\ref{alg:fock-dv-dipe} presents the resulting end-to-end
CV-DIPE protocol. It allocates $\ve/2$ to the certified Fock truncation
bias and $\ve/2$ to the statistical error of the DV-DIPE estimator.
We prove the following worst-case sample complexity guarantee.

\begin{algorithm}
\caption{End-to-end CV-DIPE via coordinated Haar measurements}
\label{alg:fock-dv-dipe}
\begin{algorithmic}[1]
\REQUIRE Unknown $m$-mode states $\rho,\sigma\in\cC_{m,E}$;
energy bound $E$; accuracy $\ve\in(0,1)$ and
failure probability $\delta\in(0,1)$.
\ENSURE An estimate of $\tr(\rho\sigma)$ with additive error at most
$\ve$, except with probability $\delta$.
\STATE Set the common mode cutoff
$L=\max\{1,\lceil4mE/\ve\rceil\}$, the dimension $D_L=L^m+2$,
and $\ve_{\rm stat}=\ve/2$.
\STATE Choose the number $N$ of copies per party as in
Eq.~\eqref{eq:fock-dv-explicit-count}.
\STATE Locally apply the maps of
Eq.~\eqref{eq:fock-dv-flag-states} to obtain $N$ copies of
$\rho_L^A$ and $\sigma_L^B$.
\STATE Run the DV-DIPE protocol of Ref.~\cite{anshu2022distributed}
on these $D_L$-dimensional states with shared Haar measurements,
statistical error tolerance $\ve_{\rm stat}$, and failure probability $\delta$.
\STATE \textbf{Return} the inner product estimator $\wh s_{\rm Haar}$.
\end{algorithmic}
\end{algorithm}

\begin{proposition}
\label{prop:fock-dv-dipe}
Let $\rho,\sigma\in\cC_{m,E}$ and
$\ve,\delta\in(0,1)$. Choose the common mode cutoff
$L=\max\{1,\lceil4mE/\ve\rceil\}$. There is a universal constant
$C>0$ such that, if the number of copies per party satisfies
\begin{align}
N\geq C\max\left\{\frac{4}{\ve^2},\;
\frac{2}{\ve}\sqrt{2+\left(1+\frac{4mE}{\ve}\right)^m}
\right\}\log\frac{2}{\delta},
\label{eq:fock-dv-explicit-count}
\end{align}
then the output of Alg.~\ref{alg:fock-dv-dipe} obeys
\begin{align}
\sup_{\rho,\sigma\in\cC_{m,E}}
\bP\!\left(\left|\wh s_{\rm Haar}
-\tr(\rho\sigma)\right|>\ve\right)\leq\delta.
\end{align}
\end{proposition}

The proof is given in
Appx.~\ref{appx:proof-of-proposition-fock-dv-dipe}.
For fixed $m$ and $E\geq1/2$, the sufficient copy count in
Eq.~\eqref{eq:fock-dv-explicit-count} scales as
$\cO(\ve^{-\max\{2,1+m/2\}}\log(2/\delta))$ copies per party.

\summary{Implementing Haar measurements.}
In the DV-DIPE protocol, each laboratory must apply the same unitary
$U\in U(D_L)$ drawn from the Haar distribution before measuring in a fixed computational basis,
which comprises the $d_L$ retained multimode Fock states and the two
flag states. The Reck and Clements decompositions implement arbitrary
unitaries on optical \emph{modes} using beamsplitters and phase
shifters~\cite{reck1994experimental,clements2016optimal}. Acting on the
original bosonic modes, however, these passive circuits conserve total
photon number and cannot realize a generic $U\in U(D_L)$ that mixes
different sectors of total photon number. A random Haar interferometer does not implement the required measurement.

One possible implementation coherently maps the $D_L$-dimensional input
space to a single photon distributed among $D_L$ optical rails. A Reck
or Clements interferometer can then implement $U$, followed by
photon detection that resolves the rails. This construction uses $D_L$ rails and
$\cO(D_L^2)$ interferometric elements. Transferring arbitrary
superpositions in the retained Fock sector to the single-photon encoding
and preparing the overflow flags require resources beyond passive
optics, such as nonlinear control or ancillary photons with detection
and feedforward~\cite{knill2001scheme,yu2026nonlinearity}. Although the
interferometer implementing $U$ need not contain non-Gaussian unitaries,
the full measurement is experimentally challenging because it requires the
cutoff and flag operations, coherent encoding, a large stable
interferometer, and resolved photon detection.
The obtained sample complexity does not account for these implementation costs.

\section{CV-DIPE via Independent Homodyne Shadows}
\label{sec:cv-dipe-shadows}

Continuous-variable classical shadows assign an operator snapshot to each
randomized measurement outcome, allowing properties of a bosonic state to be
estimated from its classical measurement records~\cite{gandhari2024precision,becker2024classical,iosue2024continuous}.
Discrete-variable classical shadow methods have already been applied to
DIPE~\cite{anshu2022distributed,zhu2022cross,huang2026worst}, 
motivating their continuous-variable counterparts for CV-DIPE. 
We study independent homodyne shadow estimators for
arbitrary, possibly entangled and non-Gaussian, $m$-mode inputs in the energy-constrained
class $\cC_{m,E}$ of Eq.~\eqref{eq:energy-bounded-state}. 
Each laboratory measures its own copies with independently chosen local
oscillator phases and reconstructs snapshots of finite rank in Fock space.
The energy constraint certifies the cutoff bias, while the snapshot
fluctuations determine the required copy count.

\summary{Homodyne snapshots and cutoff bias.}
For a cutoff $L$, use the projector $P_L$ and tail bound $q_L$ of
Eq.~\eqref{eq:fock-energy-cutoff}. On each copy and mode $j$, draw
$\theta_j$ independently and uniformly from $[0,\pi)$, measure the
quadrature $\hat x_{\theta_j}:=(a_je^{-i\theta_j}
+a_j^\dagger e^{i\theta_j})/\sqrt{2}$, and record its outcome
$x_j\in\bR$. We use the quadrature, pattern function, and cutoff
conventions of Eqs.~(8) and (11)--(13) of
Ref.~\cite{gandhari2024precision}.
Let $f_{rs}$ be the pattern function of Eq.~(12) of that reference.
For $0\leq r,s<L$, define the one-mode snapshot by
\begin{align}
\bra{r}X_{L,j}^{\rm hom}(x_j,\theta_j)\ket{s}
&:=e^{i(r-s)\theta_j}f_{rs}(x_j),
\label{eq:homodyne-snapshot}
\end{align}
with all other matrix elements zero. The phase density
$1/\pi$ belongs to the random draw of $\theta_j$ and is not an
additional factor in the snapshot.
The multimode snapshot is $\hat\tau^{\rm hom}=\bigotimes_{j=1}^m X_{L,j}^{\rm hom}$.
The local reconstruction identities give
\begin{align}
\bE_\tau[\hat\tau^{\rm hom}]=P_L\tau P_L
\label{eq:homodyne-projected-mean}
\end{align}
for every $\tau\in\cC_{m,E}$, including states entangled across modes.
The measurement outcomes within one laboratory may be correlated; the
product form of the snapshot does not assume a product input state.
Eq.~\eqref{eq:fock-dv-bias} bounds the difference between the exact
inner product and the inner product of the expected snapshots by $2q_L$.

Alice and Bob collect $N$ independent snapshots
$\hat\rho_1^{\rm hom},\ldots,\hat\rho_N^{\rm hom}$ and
$\hat\sigma_1^{\rm hom},\ldots,\hat\sigma_N^{\rm hom}$, respectively.
They use all cross-party pairs to construct the estimator
\begin{align}
\wh s_{\rm hom}
:=\frac1{N^2}\sum_{t,u=1}^{N}
\tr\!\left(\hat\rho_t^{\rm hom}\hat\sigma_u^{\rm hom}\right).
\label{eq:homodyne-shadow-estimator}
\end{align}
The measurement phases need not be shared. Independence between the
laboratories and Eq.~\eqref{eq:homodyne-projected-mean} give
$\bE[\wh s_{\rm hom}]=\tr(P_L\rho P_L\,P_L\sigma P_L)$.

For a real random variable $Z$, let
$\Var[Z]:=\bE[(Z-\bE[Z])^2]$ be its variance.
Using the pointwise bounds for one mode from
Ref.~\cite{gandhari2024precision}, we show in
Appx.~\ref{appx:cv-dipe-shadows} that, for constants
$C_1,C_2>0$ independent of $L$,
\begin{align}
\Var[\wh s_{\rm hom}]
\leq\frac{2C_1^mL^{7m/3}}{N}
+\frac{C_2^{2m}L^{6m}}{N^2}.
\label{eq:homodyne-shadow-variance}
\end{align}
Alg.~\ref{alg:homodyne-cv-dipe} specifies the end-to-end protocol.
It assigns $\ve/2$ to the Fock cutoff bias and $\ve/2$ to statistical
error. We prove the following worst-case sample complexity guarantee for the exact inner product.

\begin{algorithm}
\caption{CV-DIPE via independent homodyne shadows}
\label{alg:homodyne-cv-dipe}
\begin{algorithmic}[1]
\REQUIRE Unknown $m$-mode states $\rho,\sigma\in\cC_{m,E}$;
energy bound $E$; accuracy $\ve\in(0,1)$ and
failure probability $\delta\in(0,1)$.
\ENSURE An estimate of $\tr(\rho\sigma)$ with additive error at most
$\ve$, except with probability $\delta$.
\STATE Set the common mode cutoff
$L=\max\{1,\lceil4mE/\ve\rceil\}$ and the statistical tolerance
$\ve_{\rm stat}=\ve/2$.
\STATE Choose the number $N$ of copies per party as in
Eq.~\eqref{eq:homodyne-shadow-dipe}.
\FOR{$t=1,\ldots,N$}
\STATE Alice measures a fresh copy of $\rho$ with product homodyne
detection and constructs $\hat\rho_t^{\rm hom}$.
\STATE Bob independently measures a fresh copy of $\sigma$ and
constructs $\hat\sigma_t^{\rm hom}$.
\ENDFOR
\STATE \textbf{Return} $\wh s_{\rm hom}$ from
Eq.~\eqref{eq:homodyne-shadow-estimator}.
\end{algorithmic}
\end{algorithm}

\begin{table*}[t]
\centering
\setlength{\tabcolsep}{0pt}
\setlength\heavyrulewidth{0.3ex}
\renewcommand{\arraystretch}{1.5}
\begin{tabular}{@{}p{0.21\textwidth}p{0.34\textwidth}p{0.24\textwidth}p{0.21\textwidth}@{}}
\toprule
\textbf{Protocol} & \textbf{Sample complexity per party}
& \textbf{Leading term scaling}
& \textbf{Experimental feasibility} \\
\midrule
\makecell[l]{Coordinated\\displaced parity\\(Alg.~\ref{alg:cv-dipe})}
& $\displaystyle\begin{aligned}
N&\geq\frac{8}{\ve^2}\left(\frac{8eE}{\ve}\right)^{2m}
\log\frac{2}{\delta}
\end{aligned}$
& $\displaystyle\mathcal O_m\!\left(\left(\frac{eE}{\ve}\right)^{2m}\right)$
& Moderate \\
\addlinespace[0.3em]
\makecell[l]{Coordinated\\Haar measurements\\(Alg.~\ref{alg:fock-dv-dipe})}
& $\displaystyle\begin{aligned}
N&\geq C\max\left\{\frac{4}{\ve^2},\frac{2\sqrt{2+D_4}}{\ve}\right\}
\log\frac{2}{\delta}
\end{aligned}$
& $\displaystyle\mathcal O_m\!\left(\left(\frac{eE}{\ve}\right)^{m/2}\right)$
& Challenging \\
\addlinespace[0.3em]
\makecell[l]{Independent\\homodyne shadows\\(Alg.~\ref{alg:homodyne-cv-dipe})}
& $\displaystyle\begin{aligned}
N\geq\left[
1+\frac{8C_1^mD_4^{7/3}}{\ve^2}
+\frac{2C_2^mD_4^3}{\ve}\right]\frac{2}{\delta}
\end{aligned}$
& $\displaystyle\mathcal O_m\!\left(\left(\frac{eE}{\ve}\right)^{3m}\right)$
& Feasible \\
\bottomrule
\end{tabular}
\caption{\raggedright
\textbf{Worst-case sample complexity bounds for three CV-DIPE protocols.}
The sufficient bounds apply to arbitrary $\rho,\sigma\in\cC_{m,E}$
at accuracy $\ve\in(0,1)$ and failure
probability $\delta\in(0,1)$.
For compactness, $D_4:=(1+4mE/\ve)^m$.
The rows restate Eqs.~\eqref{eq:energy-copy-complexity},
\eqref{eq:fock-dv-explicit-count}, and
\eqref{eq:homodyne-shadow-dipe}, in row order.
The constants $C,C_1,C_2$ are as in
Eqs.~\eqref{eq:fock-dv-explicit-count} and
\eqref{eq:homodyne-shadow-dipe}. 
The $\mathcal O_m$ entries summarize only the powers of $E$ and
$1/\ve$ proportional to $m$ in the leading terms at large $m$.
They suppress all other factors, including additional accuracy and
confidence dependence; the preceding column gives the full sufficient
bounds.
The feasibility ratings refer to the optical operations discussed in
Sec.~\ref{subsec:experimental-feasibility}.}
\label{tab:cv-dipe-protocol-comparison}
\end{table*}

\begin{proposition}
\label{prop:homodyne-shadow-dipe}
Let $\rho,\sigma\in\cC_{m,E}$ and $\ve,\delta\in(0,1)$.
Choose the common mode cutoff
$L=\max\{1,\lceil4mE/\ve\rceil\}$.
If the number of copies per party satisfies
\begin{align}
N\geq\left\lceil\left[
\begin{aligned}
&1+\frac{8C_1^m}{\ve^2}
\left(1+\frac{4mE}{\ve}\right)^{7m/3}\\
&+\frac{2C_2^m}{\ve}
\left(1+\frac{4mE}{\ve}\right)^{3m}
\end{aligned}\right]\frac{2}{\delta}\right\rceil,
\label{eq:homodyne-shadow-dipe}
\end{align}
then the output of Alg.~\ref{alg:homodyne-cv-dipe} obeys
\begin{align}
\sup_{\rho,\sigma\in\cC_{m,E}}
\bP\!\left(\left|\wh s_{\rm hom}-\tr(\rho\sigma)\right|>\ve\right)
\leq\delta.
\end{align}
\end{proposition}
The proof is given in
Appx.~\ref{appx:cv-dipe-shadows}.
A standard median of means construction~\cite{elben2022randomized} using independent batches of
this estimator can further reduce the confidence dependence to
$\ln(1/\delta)$ without sharing homodyne phases between laboratories. The copy count
above follows from homodyne snapshot norms specialized for inner product
estimation; the tomography theorem of Ref.~\cite{gandhari2024precision}
uses a different error criterion.

\section{Comparison of CV-DIPE Protocols}
\label{sec:comparison}

For arbitrary, possibly entangled $\rho,\sigma\in\cC_{m,E}$, coordinated
displaced parity (Alg.~\ref{alg:cv-dipe}), coordinated Haar measurements
(Alg.~\ref{alg:fock-dv-dipe}), and independent homodyne shadows
(Alg.~\ref{alg:homodyne-cv-dipe}) each estimate $\tr(\rho\sigma)$ to
additive error $\ve\in(0,1)$ with failure probability at most
$\delta\in(0,1)$ using local measurements and classical communication.
Their worst-case sample complexity bounds are given by
Cor.~\ref{cor:energy-end-to-end}, Prop.~\ref{prop:fock-dv-dipe}, and
Prop.~\ref{prop:homodyne-shadow-dipe}, respectively, and compared in
Tab.~\ref{tab:cv-dipe-protocol-comparison}.
We compare these bounds with the required measurements to draw qualitative
observations that may help select a suitable CV-DIPE protocol.

\summary{Haar measurements have milder leading scaling in the sufficient bounds.}
The Haar protocol measures the retained multimode Fock space jointly
within each laboratory, whereas the parity and homodyne protocols use
modewise quantum measurements. None requires a joint measurement across
the two laboratories. In the leading terms of
Tab.~\ref{tab:cv-dipe-protocol-comparison}, the mode-dependent powers
of both $eE$ and $1/\ve$ are $m/2$ for Haar, $2m$ for parity, and
$3m$ for homodyne. This suggests a possible statistical
benefit from global measurements, at the cost of a challenging
multimode Haar implementation. We emphasize that these are sufficient bounds rather than
matching lower bounds; the protocols also differ in their estimators
and, for homodyne, in coordination. The finite-dimensional optimality
result of Ref.~\cite{anshu2022distributed} does not establish CV
optimality under the total energy constraint.

\summary{Parity and homodyne have different bounds for local measurements.}
Both protocols use modewise quantum measurements, but the parity
protocol applies shared random displacements while homodyne shadows
choose their measurement phases independently. The mode-dependent
powers of both $eE$ and $1/\ve$ are $2m$ for parity and $3m$ for homodyne. 
Note that the energy and accuracy bounds do not isolate the effect of shared
randomness, since the observables, truncations, estimators, and
concentration analyses differ.

\subsection{Experimental feasibility}
\label{subsec:experimental-feasibility}

The ratings in Tab.~\ref{tab:cv-dipe-protocol-comparison} compare the
optical operations required at each laboratory. The two coordinated
protocols share measurement settings between laboratories, whereas
homodyne shadows use independent settings.

\emph{Feasible} denotes measurements based on standard optical
components. Independent homodyne shadows require a balanced homodyne
measurement on each mode with a chosen local oscillator phase. The Fock
cutoff is imposed when processing the measurement records into snapshots;
it does not require a physical projection onto the Fock subspace.

\emph{Moderate} denotes an additional demanding readout. Coordinated
displaced parity uses optical displacements at shared points in phase space
and measures the parity of the total photon number in each laboratory.
The displacements are standard, but detection that resolves parity, for
example through photon counting that resolves the occupation of each mode,
becomes more demanding as the number of modes and occupied Fock levels
increase.

\emph{Challenging} denotes coherent operations beyond these local
measurements. Coordinated Haar measurements require a cutoff channel
with orthogonal overflow flags and a common basis drawn from the Haar distribution 
over the retained multimode Fock space augmented by the two flag states. 
Passive interferometers acting on the
original modes cannot implement a generic basis that mixes different
total photon number sectors. The optical rail construction of
Sec.~\ref{sec:fock-dv-dipe} additionally requires coherent encoding,
a large interferometer, and resolved detection. These hardware costs
are absent from the worst-case sample complexity bounds in
Tab.~\ref{tab:cv-dipe-protocol-comparison}.

\section{Conclusions}
\label{sec:conclusion}

We have developed three CV-DIPE protocols for cross-platform
verification of unknown, separated bosonic states under a known energy
constraint. One uses coordinated displaced parity, another coordinated
Haar measurements, and the third independent homodyne shadows. For arbitrary, possibly entangled
multimode inputs, we have established certified truncation errors and
worst-case sample complexity guarantees for the exact inner product. We
have also shown that no fixed region of finite phase space volume
controls truncation bias uniformly over unrestricted states. 
The energy constraint bounds the unobserved tails, allowing finite local
measurements to provide certified comparisons of infinite-dimensional
states. The sufficient bounds suggest a statistical benefit from global
Haar measurements, whose implementation is challenging.

An important open question is the optimal worst-case copy complexity for
energy-constrained CV-DIPE, particularly whether shared randomness gives
a provable advantage when local measurement resources are held fixed.
Extending the guarantees to loss and detector noise is another step
toward testing these tradeoffs on optical devices.

\section*{Acknowledgements}

This work was supported by 
the Quantum Science and Technology-National Science and Technology Major Project (Grant No. 2025ZD0300300),
the National Natural Science Foundation of China (Grant No. 12504584), and
the Innovation Research Foundation of NUDT (Grant No. 202501-YJRC-ZZ-004).

\textbf{Note added.} We became aware of parallel work by Tang
\emph{et al.}~\cite{tang2026distributed}, who use shared homodyne phases
and a characteristic function estimator for CV-DIPE. Our coordinated
displaced parity protocol likewise combines shared settings with
certified phase space truncation, but estimates a Wigner function
integral from parity outcomes at common displacement points. We also
develop two new CV-DIPE protocols not considered there.
%

\makeatletter
\newcommand{\appendixtitle}[1]{\gdef\@title{#1}}
\makeatother

\makeatletter%
\newcommand{\appendixmaketitle}{%
\begin{center}%
\vspace{0.4in}%
{\large \@title \par}%
\end{center}%
\par%
}%
\makeatother%

\makeatletter
\newcommand{\appendixtableofcontents}{%
\begingroup
\setcounter{tocdepth}{4}%
\@starttoc{atoc}%
\endgroup}
\makeatother

\newcommand{\appendixtocdivider}{%
\par\medskip
\noindent\hbox to \linewidth{%
\leaders\hbox{\rule[0.6ex]{1pt}{0.4pt}}\hfill
\hspace{0.8em}\textsc{Table of Contents}\hspace{0.8em}%
\leaders\hbox{\rule[0.6ex]{1pt}{0.4pt}}\hfill}%
\par}

\newcommand{\appsection}[1]{%
\section{#1}%
\addcontentsline{atoc}{section}{\protect\numberline{\thesection}#1}}

\newcommand{\appsubsection}[1]{%
\subsection{#1}%
\addcontentsline{atoc}{subsection}{\protect\numberline{\thesubsection}#1}}

\setcounter{secnumdepth}{4}
\appendix
\widetext
\newpage

\appendixtitle{\bf
Supplemental Material for\\``\thetitle''}
\appendixmaketitle
\vspace{0.1in}

\appendixtocdivider
{%
\appendixtableofcontents%
}%

\appsection{CV-DIPE via Coordinated Displaced Parity}
\label{appx:cv-dipe-parity}

\appsubsection{Proof of Prop.~\ref{prop:impossible}}\label{appx:proof-of-proposition-impossible}

Because $0\leq\cT_\Omega(\rho,\rho)\leq\tr(\rho^2)\leq1$, the displayed supremum is at most $1$. To prove the matching lower bound, let $\ket{\beta}$ be the multimode coherent state with displacement $\beta\in\bC^m$ and set $\rho_\beta=\ketbra{\beta}$. For a vector $z\in\bC^m$, let $\norm{z}{2}:=(\sum_{j=1}^m|z_j|^2)^{1/2}$ denote its Euclidean norm. The Wigner functions of these states are Gaussian translates,
\begin{align}
W_{\rho_\beta}(\boldsymbol{\alpha})=\left(\frac{2}{\pi}\right)^m e^{-2\norm{\boldsymbol{\alpha}-\beta}{2}^2},
\end{align}
so $\tr(\rho_\beta^2)=1$ and
\begin{align}
\cT_\Omega(\rho_\beta,\rho_\beta)=\pi^m\int_\Omega W_{\rho_\beta}(\boldsymbol{\alpha})^2\,d\boldsymbol{\alpha}
=\left(\frac{4}{\pi}\right)^m \int_\Omega e^{-4\norm{\boldsymbol{\alpha}-\beta}{2}^2}\,d\boldsymbol{\alpha}.
\end{align}
Since $\Omega$ has finite volume and the Gaussian is integrable, this integral tends to $0$ as $\norm{\beta}{2}\to\infty$. Hence
\begin{align}
\lim_{\norm{\beta}{2}\to\infty}\left|1-\cT_\Omega(\rho_\beta,\rho_\beta)\right|=1,
\end{align}
which proves the claim.


\appsubsection{Energy tail bound and proof of Prop.~\ref{prop:energy-truncation-bias}}\label{appx:energy-tail-bias}

For an operator $A$, let $\norm{A}{2}:=\sqrt{\tr(A^\dagger A)}$ denote its Hilbert--Schmidt norm whenever it is finite. Let $\hat\nu_j:=a_j^\dagger a_j$ be the photon number operator of mode $j$. For each mode, the Wigner--Weyl correspondence and the Plancherel identity give
\begin{align}
\pi^m\int_{\bC^m}|\alpha_j|^2|W_\tau(\boldsymbol{\alpha})|^2\,d\boldsymbol{\alpha}
&=\frac{1}{4}\norm{a_j\tau+\tau a_j}{2}^2\\
&\leq\frac{1}{2}\left(\norm{a_j\tau}{2}^2+\norm{\tau a_j}{2}^2\right)\\
&=\frac{1}{2}\tr\!\left(\hat \nu_j\tau^2+(\hat \nu_j+\1)\tau^2\right)\\
&\leq \tr(\hat \nu_j\tau)+\frac{1}{2},
\label{eq:wigner-second-moment}
\end{align}
where the last line uses $0\leq\tau^2\leq\tau$. Summing over the modes and using
Eq.~\eqref{eq:energy-bounded-state} gives
\begin{align}
\pi^m\int_{\bC^m}\sum_{j=1}^m|\alpha_j|^2
|W_\tau(\boldsymbol\alpha)|^2\,d\boldsymbol\alpha
&\leq \tr(\hat\nu_{\rm tot}\tau)+\frac m2
\leq mE.
\end{align}
Outside $B_R$, the squared radius in phase space exceeds $R^2$, so
$\pi^m\norm{W_\tau\mathbbm{1}_{B_R^c}}{2}^2\leq mE/R^2$.
Taking square roots and the supremum over $\tau\in\cC_{m,E}$
proves Eq.~\eqref{eq:energy-wigner-tail}.

For $\rho,\sigma\in\cC_{m,E}$, the Wigner inner product identity and
Cauchy--Schwarz give
\begin{align}
\left|\tr(\rho\sigma)-\cT_{B_R}(\rho,\sigma)\right|
&=\pi^m\left|\int_{B_R^c}
W_\rho(\boldsymbol\alpha)W_\sigma(\boldsymbol\alpha)
\,d\boldsymbol\alpha\right|\\
&\leq\pi^m
\norm{W_\rho\mathbbm{1}_{B_R^c}}{2}
\norm{W_\sigma\mathbbm{1}_{B_R^c}}{2}\\
&\leq\pi^m\eta_{B_R}^2
=\frac{mE}{R^2}.
\end{align}
Taking the supremum over both inputs proves
Prop.~\ref{prop:energy-truncation-bias}.


\appsubsection{Proof of Thm.~\ref{thm:hoeffding}}\label{appx:proof-of-theorem-hoeffding}

The random variables
\begin{align}
Z_t:=\left(\frac{4}{\pi}\right)^m |B_R|X_tY_t
\end{align}
are i.i.d., satisfy $\bE[Z_t]=\cT_{B_R}(\rho,\sigma)$ by Eq.~\eqref{eq:unbiased-truncated}, and obey $|Z_t|\le c_{B_R}$ with
\begin{align}
c_{B_R}:=\left(\frac{4}{\pi}\right)^m|B_R|.
\end{align}
Because each $Z_t$ lies in the interval $[-c_{B_R},c_{B_R}]$, Hoeffding's inequality gives
\begin{align}
\bP\left(\left|\frac{1}{N}\sum_{t=1}^N Z_t-\bE[Z_1]\right|\ge \ve_{\rm stat}\right)
\le 2\exp\!\left(-\frac{N\ve_{\rm stat}^2}{2c_{B_R}^2}\right).
\end{align}
The bound is independent of $\rho$ and $\sigma$. Taking the supremum over
all states with $m$ modes and requiring the bound to be at most $\delta$
proves the claim.



\appsection{CV-DIPE via Coordinated Haar Measurements}
\label{appx:cv-dipe-haar}

\appsubsection{Fock cutoff inner product bound}
\label{appx:fock-cutoff-overlap}


We prove that the bias bound in Eq.~\eqref{eq:fock-dv-bias} holds uniformly over the class.
Write $P=P_L$, $Q=\1-P$, and $q_\tau=\tr(Q\tau)$.
An occupation outcome outside $P$ has total photon number at least $L$.
Markov's inequality therefore gives $q_\tau\leq q_L$ for every
$\tau\in\cC_{m,E}$. In the $P\oplus Q$ block decomposition,
positivity of $\tau$ implies
\begin{align}
\norm{\tau-P\tau P}{2}^{2}
=2\norm{P\tau Q}{2}^{2}+\norm{Q\tau Q}{2}^{2}
\leq2(1-q_\tau)q_\tau+q_\tau^2
\leq2q_L.
\label{eq:shadow-energy-fock-hs-tail}
\end{align}
The map $A\mapsto PAP$ is an orthogonal projection on the
Hilbert--Schmidt operator space. To see explicitly why the cross terms
vanish, set $\Delta_\tau:=\tau-P\tau P$. Since $P^2=P$,
$P\Delta_\tau P=0$. Cyclicity of the trace and the expansions
$\rho=P\rho P+\Delta_\rho$ and
$\sigma=P\sigma P+\Delta_\sigma$ then give, for
$\rho,\sigma\in\cC_{m,E}$,
\begin{align}
\tr(\Delta_\rho P\sigma P)
&=\tr(P\Delta_\rho P\,\sigma)=0,\\
\tr(P\rho P\,\Delta_\sigma)
&=\tr(\rho P\Delta_\sigma P)=0,\\
\tr(\rho\sigma) 
&= \tr((P\rho P+\Delta_\rho)(P\sigma P+\Delta_\sigma))
= \tr(P\rho P P\sigma P) + \tr((\rho - P\rho P)(\sigma - P\sigma P)).
\end{align}
Then we have
\begin{align}
\left|\tr(\rho\sigma)-\tr(P\rho P\,P\sigma P)\right|
&=\left|\tr\!\left[(\rho-P\rho P)(\sigma-P\sigma P)\right]\right|\\
&\leq\norm{\rho-P\rho P}{2}\norm{\sigma-P\sigma P}{2}\\
&\leq\sqrt{2q_\rho}\sqrt{2q_\sigma}
=2\sqrt{q_\rho q_\sigma}\leq2q_L.
\end{align}
The first inequality is Hilbert--Schmidt Cauchy--Schwarz, the second
uses Eq.~\eqref{eq:shadow-energy-fock-hs-tail}, and the last uses
$q_\rho,q_\sigma\leq q_L$.


\appsubsection{Proof of Prop.~\ref{prop:fock-dv-dipe}}
\label{appx:proof-of-proposition-fock-dv-dipe}

For the stated cutoff, $L\geq4mE/\ve$, so $q_L\leq\ve/4$.
Eqs.~\eqref{eq:fock-dv-bias} and~\eqref{eq:fock-dv-flag-overlap}
therefore give
$|\tr(\rho\sigma)-\tr(\rho_L^A\sigma_L^B)|\leq\ve/2$
uniformly over $\rho,\sigma\in\cC_{m,E}$.
The finite-dimensional DV-DIPE guarantee of
Ref.~\cite{anshu2022distributed}, applied in dimension $D_L$ with
statistical tolerance $\ve/2$, bounds the failure probability by
$\delta$ when the number of copies per party is at least
$C_0\log(2/\delta)\max\{4/\ve^2,2\sqrt{D_L}/\ve\}$ for a universal
constant $C_0>0$. Since $L\leq1+4mE/\ve$, we have
$D_L\leq2+(1+4mE/\ve)^m$. Thus the count in
Eq.~\eqref{eq:fock-dv-explicit-count}, with a sufficiently large
universal $C$, meets the finite-dimensional requirement. The triangle
inequality then gives the claimed error and failure probability.


\appsection{Proof of Prop.~\ref{prop:homodyne-shadow-dipe}}
\label{appx:cv-dipe-shadows}

For an input pair, let $X$ and $Y$ denote independent homodyne
snapshots with means $\mu_\rho=P_L\rho P_L$ and
$\mu_\sigma=P_L\sigma P_L$ from
Eq.~\eqref{eq:homodyne-projected-mean}. Set
$X_0=X-\mu_\rho$ and $Y_0=Y-\mu_\sigma$.
Define $v_\rho:=\bE[\tr(X_0\mu_\sigma)^2]$,
$v_\sigma:=\bE[\tr(\mu_\rho Y_0)^2]$, and
$q_{\rm hom}:=\bE[\tr(X_0Y_0)^2]$, where the last expectation is over
independent snapshots.
The estimator in Eq.~\eqref{eq:homodyne-shadow-estimator} is the inner
product of the two empirical mean snapshots. Independence and
centering give
\begin{align}
\bE[\wh s_{\rm hom}]&=\tr(\mu_\rho\mu_\sigma),\\
\Var[\wh s_{\rm hom}]
&=\frac{v_\rho+v_\sigma}{N}+\frac{q_{\rm hom}}{N^2}.
\end{align}
For an operator $T$, let $\norm{T}{\infty}$ denote its operator norm.
The one-mode bounds appear in the proof of Theorem 1 of
Ref.~\cite{gandhari2024precision}, immediately before its Eq.~(14).
For the snapshot defined by Eq.~\eqref{eq:homodyne-snapshot}, they give
$\norm{X_{L,j}^{\rm hom}(x_j,\theta_j)}{\infty}^{2}
\leq C_1L^{7/3}$ and
$\norm{X_{L,j}^{\rm hom}(x_j,\theta_j)}{2}^{2}\leq C_2L^3$.
The pattern function bound used there takes the supremum over outcomes
$x_j$. It also bounds the squared Hilbert--Schmidt norm pointwise by a
constant times $L^{7/3}$, which is at most a constant times $L^3$ for
$L\geq1$; we enlarge $C_2$ if needed. The phase factors in
Eq.~\eqref{eq:homodyne-snapshot} give a
diagonal unitary conjugation, so both norms are independent of
$\theta_j$. Thus $C_1,C_2$ can be chosen independently of $L$, $x_j$,
and $\theta_j$.
The multimode snapshots are tensor products, so
$\norm{X}{\infty}^{2}\leq C_1^mL^{7m/3}$ and
$\norm{X}{2}^{2}\leq C_2^mL^{3m}$, and likewise for $Y$.
These pointwise bounds hold even when the outcomes are correlated
across modes by an entangled input.
The projected means are positive with trace at most one. Since
centering cannot increase scalar variance,
$v_\rho,v_\sigma\leq C_1^mL^{7m/3}$.
Also,
$\bE[\norm{X_0}{2}^{2}]\leq C_2^mL^{3m}$ and likewise for $Y_0$.
Hilbert--Schmidt Cauchy--Schwarz and independence give
$q_{\rm hom}\leq C_2^{2m}L^{6m}$, proving
Eq.~\eqref{eq:homodyne-shadow-variance}.

The stated cutoff satisfies $q_L\leq\ve/4$, so
Eq.~\eqref{eq:fock-dv-bias} bounds the bias by $\ve/2$.
Since $L\leq1+4mE/\ve$, the copy count in
Eq.~\eqref{eq:homodyne-shadow-dipe} is at least
$16C_1^mL^{7m/3}/(\delta\ve^2)$ and
$4C_2^mL^{3m}/(\delta\ve)$.
Eq.~\eqref{eq:homodyne-shadow-variance} is therefore at most
$(\delta/8+\delta^2/16)\ve^2<\delta\ve^2/4$.
Chebyshev's inequality bounds the probability of statistical error
greater than $\ve/2$ by $\delta$. The triangle inequality gives the
claimed guarantee for the exact inner product, uniformly over
$\cC_{m,E}$.


\begin{thebibliography}{38}%
\makeatletter
\providecommand \@ifxundefined [1]{%
 \@ifx{#1\undefined}
}%
\providecommand \@ifnum [1]{%
 \ifnum #1\expandafter \@firstoftwo
 \else \expandafter \@secondoftwo
 \fi
}%
\providecommand \@ifx [1]{%
 \ifx #1\expandafter \@firstoftwo
 \else \expandafter \@secondoftwo
 \fi
}%
\providecommand \natexlab [1]{#1}%
\providecommand \enquote  [1]{``#1''}%
\providecommand \bibnamefont  [1]{#1}%
\providecommand \bibfnamefont [1]{#1}%
\providecommand \citenamefont [1]{#1}%
\providecommand \href@noop [0]{\@secondoftwo}%
\providecommand \href [0]{\begingroup \@sanitize@url \@href}%
\providecommand \@href[1]{\@@startlink{#1}\@@href}%
\providecommand \@@href[1]{\endgroup#1\@@endlink}%
\providecommand \@sanitize@url [0]{\catcode `\\12\catcode `\$12\catcode
  `\&12\catcode `\#12\catcode `\^12\catcode `\_12\catcode `\%12\relax}%
\providecommand \@@startlink[1]{}%
\providecommand \@@endlink[0]{}%
\providecommand \url  [0]{\begingroup\@sanitize@url \@url }%
\providecommand \@url [1]{\endgroup\@href {#1}{\urlprefix }}%
\providecommand \urlprefix  [0]{URL }%
\providecommand \Eprint [0]{\href }%
\providecommand \doibase [0]{https://doi.org/}%
\providecommand \selectlanguage [0]{\@gobble}%
\providecommand \bibinfo  [0]{\@secondoftwo}%
\providecommand \bibfield  [0]{\@secondoftwo}%
\providecommand \translation [1]{[#1]}%
\providecommand \BibitemOpen [0]{}%
\providecommand \bibitemStop [0]{}%
\providecommand \bibitemNoStop [0]{.\EOS\space}%
\providecommand \EOS [0]{\spacefactor3000\relax}%
\providecommand \BibitemShut  [1]{\csname bibitem#1\endcsname}%
\let\auto@bib@innerbib\@empty
\bibitem [{\citenamefont {Eisert}\ \emph {et~al.}(2020)\citenamefont {Eisert},
  \citenamefont {Hangleiter}, \citenamefont {Walk}, \citenamefont {Roth},
  \citenamefont {Markham}, \citenamefont {Parekh}, \citenamefont {Chabaud},\
  and\ \citenamefont {Kashefi}}]{eisert2020quantuma}%
  \BibitemOpen
  \bibfield  {author} {\bibinfo {author} {\bibfnamefont {J.}~\bibnamefont
  {Eisert}}, \bibinfo {author} {\bibfnamefont {D.}~\bibnamefont {Hangleiter}},
  \bibinfo {author} {\bibfnamefont {N.}~\bibnamefont {Walk}}, \bibinfo {author}
  {\bibfnamefont {I.}~\bibnamefont {Roth}}, \bibinfo {author} {\bibfnamefont
  {D.}~\bibnamefont {Markham}}, \bibinfo {author} {\bibfnamefont
  {R.}~\bibnamefont {Parekh}}, \bibinfo {author} {\bibfnamefont
  {U.}~\bibnamefont {Chabaud}},\ and\ \bibinfo {author} {\bibfnamefont
  {E.}~\bibnamefont {Kashefi}},\ }\href
  {https://doi.org/10.1038/s42254-020-0186-4} {\bibfield  {journal} {\bibinfo
  {journal} {Nature Reviews Physics}\ }\textbf {\bibinfo {volume} {2}},\
  \bibinfo {pages} {382} (\bibinfo {year} {2020})}\BibitemShut {NoStop}%
\bibitem [{\citenamefont {Elben}\ \emph {et~al.}(2020)\citenamefont {Elben},
  \citenamefont {Vermersch}, \citenamefont {{van Bijnen}}, \citenamefont
  {Kokail}, \citenamefont {Brydges}, \citenamefont {Maier}, \citenamefont
  {Joshi}, \citenamefont {Blatt}, \citenamefont {Roos},\ and\ \citenamefont
  {Zoller}}]{elben2020cross}%
  \BibitemOpen
  \bibfield  {author} {\bibinfo {author} {\bibfnamefont {A.}~\bibnamefont
  {Elben}}, \bibinfo {author} {\bibfnamefont {B.}~\bibnamefont {Vermersch}},
  \bibinfo {author} {\bibfnamefont {R.}~\bibnamefont {{van Bijnen}}}, \bibinfo
  {author} {\bibfnamefont {C.}~\bibnamefont {Kokail}}, \bibinfo {author}
  {\bibfnamefont {T.}~\bibnamefont {Brydges}}, \bibinfo {author} {\bibfnamefont
  {C.}~\bibnamefont {Maier}}, \bibinfo {author} {\bibfnamefont {M.~K.}\
  \bibnamefont {Joshi}}, \bibinfo {author} {\bibfnamefont {R.}~\bibnamefont
  {Blatt}}, \bibinfo {author} {\bibfnamefont {C.~F.}\ \bibnamefont {Roos}},\
  and\ \bibinfo {author} {\bibfnamefont {P.}~\bibnamefont {Zoller}},\ }\href
  {https://doi.org/10.1103/PhysRevLett.124.010504} {\bibfield  {journal}
  {\bibinfo  {journal} {Physical Review Letters}\ }\textbf {\bibinfo {volume}
  {124}},\ \bibinfo {pages} {010504} (\bibinfo {year} {2020})}\BibitemShut
  {NoStop}%
\bibitem [{\citenamefont {Zhu}\ \emph {et~al.}(2022)\citenamefont {Zhu},
  \citenamefont {Cian}, \citenamefont {Noel}, \citenamefont {Risinger},
  \citenamefont {Biswas}, \citenamefont {Egan}, \citenamefont {Zhu},
  \citenamefont {Green}, \citenamefont {Alderete}, \citenamefont {Nguyen},
  \citenamefont {Wang}, \citenamefont {Maksymov}, \citenamefont {Nam},
  \citenamefont {Cetina}, \citenamefont {Linke}, \citenamefont {Hafezi},\ and\
  \citenamefont {Monroe}}]{zhu2022cross}%
  \BibitemOpen
  \bibfield  {author} {\bibinfo {author} {\bibfnamefont {D.}~\bibnamefont
  {Zhu}}, \bibinfo {author} {\bibfnamefont {Z.~P.}\ \bibnamefont {Cian}},
  \bibinfo {author} {\bibfnamefont {C.}~\bibnamefont {Noel}}, \bibinfo {author}
  {\bibfnamefont {A.}~\bibnamefont {Risinger}}, \bibinfo {author}
  {\bibfnamefont {D.}~\bibnamefont {Biswas}}, \bibinfo {author} {\bibfnamefont
  {L.}~\bibnamefont {Egan}}, \bibinfo {author} {\bibfnamefont {Y.}~\bibnamefont
  {Zhu}}, \bibinfo {author} {\bibfnamefont {A.~M.}\ \bibnamefont {Green}},
  \bibinfo {author} {\bibfnamefont {C.~H.}\ \bibnamefont {Alderete}}, \bibinfo
  {author} {\bibfnamefont {N.~H.}\ \bibnamefont {Nguyen}}, \bibinfo {author}
  {\bibfnamefont {Q.}~\bibnamefont {Wang}}, \bibinfo {author} {\bibfnamefont
  {A.}~\bibnamefont {Maksymov}}, \bibinfo {author} {\bibfnamefont
  {Y.}~\bibnamefont {Nam}}, \bibinfo {author} {\bibfnamefont {M.}~\bibnamefont
  {Cetina}}, \bibinfo {author} {\bibfnamefont {N.~M.}\ \bibnamefont {Linke}},
  \bibinfo {author} {\bibfnamefont {M.}~\bibnamefont {Hafezi}},\ and\ \bibinfo
  {author} {\bibfnamefont {C.}~\bibnamefont {Monroe}},\ }\href
  {https://doi.org/10.1038/s41467-022-34279-5} {\bibfield  {journal} {\bibinfo
  {journal} {Nature Communications}\ }\textbf {\bibinfo {volume} {13}},\
  \bibinfo {pages} {6620} (\bibinfo {year} {2022})}\BibitemShut {NoStop}%
\bibitem [{\citenamefont {Dalton}\ \emph {et~al.}(2025)\citenamefont {Dalton},
  \citenamefont {Kn{\"o}rzer}, \citenamefont {Hoehne}, \citenamefont {Song},
  \citenamefont {Flasby}, \citenamefont {Colao~Zanuz}, \citenamefont
  {Bahrami~Panah}, \citenamefont {Besedin}, \citenamefont {Besse},\ and\
  \citenamefont {Wallraff}}]{dalton2025resourceefficient}%
  \BibitemOpen
  \bibfield  {author} {\bibinfo {author} {\bibfnamefont {K.}~\bibnamefont
  {Dalton}}, \bibinfo {author} {\bibfnamefont {J.}~\bibnamefont {Kn{\"o}rzer}},
  \bibinfo {author} {\bibfnamefont {F.}~\bibnamefont {Hoehne}}, \bibinfo
  {author} {\bibfnamefont {Y.}~\bibnamefont {Song}}, \bibinfo {author}
  {\bibfnamefont {A.}~\bibnamefont {Flasby}}, \bibinfo {author} {\bibfnamefont
  {D.}~\bibnamefont {Colao~Zanuz}}, \bibinfo {author} {\bibfnamefont
  {M.}~\bibnamefont {Bahrami~Panah}}, \bibinfo {author} {\bibfnamefont
  {I.}~\bibnamefont {Besedin}}, \bibinfo {author} {\bibfnamefont {J.-C.}\
  \bibnamefont {Besse}},\ and\ \bibinfo {author} {\bibfnamefont
  {A.}~\bibnamefont {Wallraff}},\ }\href {https://doi.org/10.1103/czph-xpzs}
  {\bibfield  {journal} {\bibinfo  {journal} {PRX Quantum}\ }\textbf {\bibinfo
  {volume} {6}},\ \bibinfo {pages} {040365} (\bibinfo {year}
  {2025})}\BibitemShut {NoStop}%
\bibitem [{\citenamefont {Wu}\ \emph {et~al.}(2025)\citenamefont {Wu},
  \citenamefont {Xie}, \citenamefont {Mi}, \citenamefont {Wu}, \citenamefont
  {Guo}, \citenamefont {Huang}, \citenamefont {Huang}, \citenamefont {Sun},
  \citenamefont {Zhang}, \citenamefont {Zhang}, \citenamefont {Qiu},
  \citenamefont {Linpeng}, \citenamefont {Tao}, \citenamefont {Chu},
  \citenamefont {Jiang}, \citenamefont {Liu}, \citenamefont {Niu},
  \citenamefont {Zhou}, \citenamefont {Du}, \citenamefont {Ren}, \citenamefont
  {Zhong}, \citenamefont {Liu},\ and\ \citenamefont {Yu}}]{wu2025state}%
  \BibitemOpen
  \bibfield  {author} {\bibinfo {author} {\bibfnamefont {B.}~\bibnamefont
  {Wu}}, \bibinfo {author} {\bibfnamefont {C.}~\bibnamefont {Xie}}, \bibinfo
  {author} {\bibfnamefont {P.}~\bibnamefont {Mi}}, \bibinfo {author}
  {\bibfnamefont {Z.}~\bibnamefont {Wu}}, \bibinfo {author} {\bibfnamefont
  {Z.}~\bibnamefont {Guo}}, \bibinfo {author} {\bibfnamefont {P.}~\bibnamefont
  {Huang}}, \bibinfo {author} {\bibfnamefont {W.}~\bibnamefont {Huang}},
  \bibinfo {author} {\bibfnamefont {X.}~\bibnamefont {Sun}}, \bibinfo {author}
  {\bibfnamefont {J.}~\bibnamefont {Zhang}}, \bibinfo {author} {\bibfnamefont
  {L.}~\bibnamefont {Zhang}}, \bibinfo {author} {\bibfnamefont
  {J.}~\bibnamefont {Qiu}}, \bibinfo {author} {\bibfnamefont {X.}~\bibnamefont
  {Linpeng}}, \bibinfo {author} {\bibfnamefont {Z.}~\bibnamefont {Tao}},
  \bibinfo {author} {\bibfnamefont {J.}~\bibnamefont {Chu}}, \bibinfo {author}
  {\bibfnamefont {J.}~\bibnamefont {Jiang}}, \bibinfo {author} {\bibfnamefont
  {S.}~\bibnamefont {Liu}}, \bibinfo {author} {\bibfnamefont {J.}~\bibnamefont
  {Niu}}, \bibinfo {author} {\bibfnamefont {Y.}~\bibnamefont {Zhou}}, \bibinfo
  {author} {\bibfnamefont {Y.}~\bibnamefont {Du}}, \bibinfo {author}
  {\bibfnamefont {W.}~\bibnamefont {Ren}}, \bibinfo {author} {\bibfnamefont
  {Y.}~\bibnamefont {Zhong}}, \bibinfo {author} {\bibfnamefont
  {T.}~\bibnamefont {Liu}},\ and\ \bibinfo {author} {\bibfnamefont
  {D.}~\bibnamefont {Yu}},\ }\href {https://doi.org/10.48550/arXiv.2506.01657}
  {\bibinfo {title} {State similarity in modular superconducting quantum
  processors with classical communications}} (\bibinfo {year} {2025}),\ \Eprint
  {https://arxiv.org/abs/2506.01657} {arXiv:2506.01657} \BibitemShut {NoStop}%
\bibitem [{\citenamefont {Zheng}\ \emph {et~al.}(2024)\citenamefont {Zheng},
  \citenamefont {Yu},\ and\ \citenamefont {Wang}}]{zheng2024crossplatform}%
  \BibitemOpen
  \bibfield  {author} {\bibinfo {author} {\bibfnamefont {C.}~\bibnamefont
  {Zheng}}, \bibinfo {author} {\bibfnamefont {X.}~\bibnamefont {Yu}},\ and\
  \bibinfo {author} {\bibfnamefont {K.}~\bibnamefont {Wang}},\ }\href
  {https://doi.org/10.1038/s41534-023-00797-3} {\bibfield  {journal} {\bibinfo
  {journal} {npj Quantum Information}\ }\textbf {\bibinfo {volume} {10}},\
  \bibinfo {pages} {1} (\bibinfo {year} {2024})}\BibitemShut {NoStop}%
\bibitem [{\citenamefont {Carrasco}\ \emph {et~al.}(2021)\citenamefont
  {Carrasco}, \citenamefont {Elben}, \citenamefont {Kokail}, \citenamefont
  {Kraus},\ and\ \citenamefont {Zoller}}]{carrasco2021theoretical}%
  \BibitemOpen
  \bibfield  {author} {\bibinfo {author} {\bibfnamefont {J.}~\bibnamefont
  {Carrasco}}, \bibinfo {author} {\bibfnamefont {A.}~\bibnamefont {Elben}},
  \bibinfo {author} {\bibfnamefont {C.}~\bibnamefont {Kokail}}, \bibinfo
  {author} {\bibfnamefont {B.}~\bibnamefont {Kraus}},\ and\ \bibinfo {author}
  {\bibfnamefont {P.}~\bibnamefont {Zoller}},\ }\href
  {https://doi.org/10.1103/prxquantum.2.010102} {\bibfield  {journal} {\bibinfo
   {journal} {PRX Quantum}\ }\textbf {\bibinfo {volume} {2}},\ \bibinfo {pages}
  {010102} (\bibinfo {year} {2021})}\BibitemShut {NoStop}%
\bibitem [{\citenamefont {Elben}\ \emph {et~al.}(2023)\citenamefont {Elben},
  \citenamefont {Flammia}, \citenamefont {Huang}, \citenamefont {Kueng},
  \citenamefont {Preskill}, \citenamefont {Vermersch},\ and\ \citenamefont
  {Zoller}}]{elben2022randomized}%
  \BibitemOpen
  \bibfield  {author} {\bibinfo {author} {\bibfnamefont {A.}~\bibnamefont
  {Elben}}, \bibinfo {author} {\bibfnamefont {S.~T.}\ \bibnamefont {Flammia}},
  \bibinfo {author} {\bibfnamefont {H.-Y.}\ \bibnamefont {Huang}}, \bibinfo
  {author} {\bibfnamefont {R.}~\bibnamefont {Kueng}}, \bibinfo {author}
  {\bibfnamefont {J.}~\bibnamefont {Preskill}}, \bibinfo {author}
  {\bibfnamefont {B.}~\bibnamefont {Vermersch}},\ and\ \bibinfo {author}
  {\bibfnamefont {P.}~\bibnamefont {Zoller}},\ }\href
  {https://doi.org/10.1038/s42254-022-00535-2} {\bibfield  {journal} {\bibinfo
  {journal} {Nature Reviews Physics}\ }\textbf {\bibinfo {volume} {5}},\
  \bibinfo {pages} {9} (\bibinfo {year} {2023})}\BibitemShut {NoStop}%
\bibitem [{\citenamefont {Jia}\ \emph {et~al.}(2025)\citenamefont {Jia},
  \citenamefont {Zhai}, \citenamefont {Zhu} \emph {et~al.}}]{jia2025microcomb}%
  \BibitemOpen
  \bibfield  {author} {\bibinfo {author} {\bibfnamefont {X.}~\bibnamefont
  {Jia}}, \bibinfo {author} {\bibfnamefont {C.}~\bibnamefont {Zhai}}, \bibinfo
  {author} {\bibfnamefont {X.}~\bibnamefont {Zhu}}, \emph {et~al.},\ }\href
  {https://doi.org/10.1038/s41586-025-08602-1} {\bibfield  {journal} {\bibinfo
  {journal} {Nature}\ }\textbf {\bibinfo {volume} {639}},\ \bibinfo {pages}
  {329} (\bibinfo {year} {2025})}\BibitemShut {NoStop}%
\bibitem [{\citenamefont {Jia}\ \emph {et~al.}(2026)\citenamefont {Jia},
  \citenamefont {You}, \citenamefont {Zhai} \emph
  {et~al.}}]{jia2026monolithic}%
  \BibitemOpen
  \bibfield  {author} {\bibinfo {author} {\bibfnamefont {X.}~\bibnamefont
  {Jia}}, \bibinfo {author} {\bibfnamefont {C.}~\bibnamefont {You}}, \bibinfo
  {author} {\bibfnamefont {C.}~\bibnamefont {Zhai}}, \emph {et~al.},\ }\href
  {https://doi.org/10.1038/s41566-026-01868-5} {\bibfield  {journal} {\bibinfo
  {journal} {Nature Photonics}\ }\textbf {\bibinfo {volume} {20}},\ \bibinfo
  {pages} {428} (\bibinfo {year} {2026})}\BibitemShut {NoStop}%
\bibitem [{\citenamefont {Konno}\ \emph {et~al.}(2024)\citenamefont {Konno},
  \citenamefont {Asavanant}, \citenamefont {Hanamura}, \citenamefont
  {Nagayoshi}, \citenamefont {Fukui}, \citenamefont {Sakaguchi}, \citenamefont
  {Ide}, \citenamefont {China}, \citenamefont {Yabuno}, \citenamefont {Miki},
  \citenamefont {Terai}, \citenamefont {Takase}, \citenamefont {Endo},
  \citenamefont {Marek}, \citenamefont {Filip}, \citenamefont {van Loock},\
  and\ \citenamefont {Furusawa}}]{konno2024logical}%
  \BibitemOpen
  \bibfield  {author} {\bibinfo {author} {\bibfnamefont {S.}~\bibnamefont
  {Konno}}, \bibinfo {author} {\bibfnamefont {W.}~\bibnamefont {Asavanant}},
  \bibinfo {author} {\bibfnamefont {F.}~\bibnamefont {Hanamura}}, \bibinfo
  {author} {\bibfnamefont {H.}~\bibnamefont {Nagayoshi}}, \bibinfo {author}
  {\bibfnamefont {K.}~\bibnamefont {Fukui}}, \bibinfo {author} {\bibfnamefont
  {A.}~\bibnamefont {Sakaguchi}}, \bibinfo {author} {\bibfnamefont
  {R.}~\bibnamefont {Ide}}, \bibinfo {author} {\bibfnamefont {F.}~\bibnamefont
  {China}}, \bibinfo {author} {\bibfnamefont {M.}~\bibnamefont {Yabuno}},
  \bibinfo {author} {\bibfnamefont {S.}~\bibnamefont {Miki}}, \bibinfo {author}
  {\bibfnamefont {H.}~\bibnamefont {Terai}}, \bibinfo {author} {\bibfnamefont
  {K.}~\bibnamefont {Takase}}, \bibinfo {author} {\bibfnamefont
  {M.}~\bibnamefont {Endo}}, \bibinfo {author} {\bibfnamefont {P.}~\bibnamefont
  {Marek}}, \bibinfo {author} {\bibfnamefont {R.}~\bibnamefont {Filip}},
  \bibinfo {author} {\bibfnamefont {P.}~\bibnamefont {van Loock}},\ and\
  \bibinfo {author} {\bibfnamefont {A.}~\bibnamefont {Furusawa}},\ }\href
  {https://doi.org/10.1126/science.adk7560} {\bibfield  {journal} {\bibinfo
  {journal} {Science}\ }\textbf {\bibinfo {volume} {383}},\ \bibinfo {pages}
  {289} (\bibinfo {year} {2024})}\BibitemShut {NoStop}%
\bibitem [{\citenamefont {Larsen}\ \emph {et~al.}(2025)\citenamefont {Larsen},
  \citenamefont {Bourassa}, \citenamefont {Kocsis} \emph
  {et~al.}}]{larsen2025gkp}%
  \BibitemOpen
  \bibfield  {author} {\bibinfo {author} {\bibfnamefont {M.~V.}\ \bibnamefont
  {Larsen}}, \bibinfo {author} {\bibfnamefont {J.~E.}\ \bibnamefont
  {Bourassa}}, \bibinfo {author} {\bibfnamefont {S.}~\bibnamefont {Kocsis}},
  \emph {et~al.},\ }\href {https://doi.org/10.1038/s41586-025-09044-5}
  {\bibfield  {journal} {\bibinfo  {journal} {Nature}\ }\textbf {\bibinfo
  {volume} {642}},\ \bibinfo {pages} {587} (\bibinfo {year}
  {2025})}\BibitemShut {NoStop}%
\bibitem [{\citenamefont {{Aghaee Rad}}\ \emph {et~al.}(2025)\citenamefont
  {{Aghaee Rad}}, \citenamefont {Ainsworth}, \citenamefont {Alexander} \emph
  {et~al.}}]{aghaee2025modular}%
  \BibitemOpen
  \bibfield  {author} {\bibinfo {author} {\bibfnamefont {H.}~\bibnamefont
  {{Aghaee Rad}}}, \bibinfo {author} {\bibfnamefont {T.}~\bibnamefont
  {Ainsworth}}, \bibinfo {author} {\bibfnamefont {R.~N.}\ \bibnamefont
  {Alexander}}, \emph {et~al.},\ }\href
  {https://doi.org/10.1038/s41586-024-08406-9} {\bibfield  {journal} {\bibinfo
  {journal} {Nature}\ }\textbf {\bibinfo {volume} {638}},\ \bibinfo {pages}
  {912} (\bibinfo {year} {2025})}\BibitemShut {NoStop}%
\bibitem [{\citenamefont {Yu}\ \emph {et~al.}(2026)\citenamefont {Yu},
  \citenamefont {Sun}, \citenamefont {Chen} \emph
  {et~al.}}]{yu2026nonlinearity}%
  \BibitemOpen
  \bibfield  {author} {\bibinfo {author} {\bibfnamefont {S.}~\bibnamefont
  {Yu}}, \bibinfo {author} {\bibfnamefont {J.}~\bibnamefont {Sun}}, \bibinfo
  {author} {\bibfnamefont {K.-C.}\ \bibnamefont {Chen}}, \emph {et~al.},\
  }\href {https://doi.org/10.1038/s41566-026-01962-8} {\bibfield  {journal}
  {\bibinfo  {journal} {Nature Photonics}\ }\textbf {\bibinfo {volume} {20}},\
  \bibinfo {pages} {1166} (\bibinfo {year} {2026})}\BibitemShut {NoStop}%
\bibitem [{\citenamefont {Anshu}\ \emph {et~al.}(2022)\citenamefont {Anshu},
  \citenamefont {Landau},\ and\ \citenamefont {Liu}}]{anshu2022distributed}%
  \BibitemOpen
  \bibfield  {author} {\bibinfo {author} {\bibfnamefont {A.}~\bibnamefont
  {Anshu}}, \bibinfo {author} {\bibfnamefont {Z.}~\bibnamefont {Landau}},\ and\
  \bibinfo {author} {\bibfnamefont {Y.}~\bibnamefont {Liu}},\ }in\ \href
  {https://doi.org/10.1145/3519935.3519974} {\emph {\bibinfo {booktitle}
  {Proceedings of the 54th Annual ACM SIGACT Symposium on Theory of
  Computing}}},\ \bibinfo {series and number} {STOC 2022}\ (\bibinfo
  {publisher} {Association for Computing Machinery},\ \bibinfo {address} {New
  York, NY, USA},\ \bibinfo {year} {2022})\ pp.\ \bibinfo {pages}
  {44--51}\BibitemShut {NoStop}%
\bibitem [{\citenamefont {Zheng}\ \emph {et~al.}(2026)\citenamefont {Zheng},
  \citenamefont {Wang}, \citenamefont {Yu}, \citenamefont {Xu},\ and\
  \citenamefont {Zhang}}]{zheng2025distributed}%
  \BibitemOpen
  \bibfield  {author} {\bibinfo {author} {\bibfnamefont {C.}~\bibnamefont
  {Zheng}}, \bibinfo {author} {\bibfnamefont {K.}~\bibnamefont {Wang}},
  \bibinfo {author} {\bibfnamefont {X.}~\bibnamefont {Yu}}, \bibinfo {author}
  {\bibfnamefont {P.}~\bibnamefont {Xu}},\ and\ \bibinfo {author}
  {\bibfnamefont {Z.}~\bibnamefont {Zhang}},\ }\href
  {https://doi.org/10.1038/s41534-026-01247-6} {\bibfield  {journal} {\bibinfo
  {journal} {npj Quantum Information}\ }\textbf {\bibinfo {volume} {12}},\
  \bibinfo {pages} {94} (\bibinfo {year} {2026})}\BibitemShut {NoStop}%
\bibitem [{\citenamefont {Huang}\ \emph {et~al.}(2026)\citenamefont {Huang},
  \citenamefont {Wang},\ and\ \citenamefont {Xu}}]{huang2026worst}%
  \BibitemOpen
  \bibfield  {author} {\bibinfo {author} {\bibfnamefont {Z.}~\bibnamefont
  {Huang}}, \bibinfo {author} {\bibfnamefont {K.}~\bibnamefont {Wang}},\ and\
  \bibinfo {author} {\bibfnamefont {P.}~\bibnamefont {Xu}},\ }\href
  {https://doi.org/10.48550/arXiv.2605.14256} {\bibinfo {title} {Worst-case
  sample complexity bounds for distributed inner product estimation with local
  randomized measurements}} (\bibinfo {year} {2026}),\ \Eprint
  {https://arxiv.org/abs/2605.14256} {arXiv:2605.14256 [quant-ph]} \BibitemShut
  {NoStop}%
\bibitem [{\citenamefont {Volkoff}\ and\ \citenamefont
  {Suba{\c{s}}{\i}}(2022)}]{volkoff2022}%
  \BibitemOpen
  \bibfield  {author} {\bibinfo {author} {\bibfnamefont {T.~J.}\ \bibnamefont
  {Volkoff}}\ and\ \bibinfo {author} {\bibfnamefont {Y.}~\bibnamefont
  {Suba{\c{s}}{\i}}},\ }\href {https://doi.org/10.22331/q-2022-09-08-800}
  {\bibfield  {journal} {\bibinfo  {journal} {Quantum}\ }\textbf {\bibinfo
  {volume} {6}},\ \bibinfo {pages} {800} (\bibinfo {year} {2022})}\BibitemShut
  {NoStop}%
\bibitem [{\citenamefont {Zhan}\ \emph {et~al.}(2025)\citenamefont {Zhan},
  \citenamefont {Wang}, \citenamefont {Mi}, \citenamefont {Xie}, \citenamefont
  {Xu}, \citenamefont {Zhang},\ and\ \citenamefont
  {Zhang}}]{zhan2025experimental}%
  \BibitemOpen
  \bibfield  {author} {\bibinfo {author} {\bibfnamefont {H.}~\bibnamefont
  {Zhan}}, \bibinfo {author} {\bibfnamefont {B.}~\bibnamefont {Wang}}, \bibinfo
  {author} {\bibfnamefont {M.}~\bibnamefont {Mi}}, \bibinfo {author}
  {\bibfnamefont {J.}~\bibnamefont {Xie}}, \bibinfo {author} {\bibfnamefont
  {L.}~\bibnamefont {Xu}}, \bibinfo {author} {\bibfnamefont {A.}~\bibnamefont
  {Zhang}},\ and\ \bibinfo {author} {\bibfnamefont {L.}~\bibnamefont {Zhang}},\
  }\href {https://doi.org/10.1038/s41377-025-01755-8} {\bibfield  {journal}
  {\bibinfo  {journal} {Light: Science \& Applications}\ }\textbf {\bibinfo
  {volume} {14}},\ \bibinfo {pages} {83} (\bibinfo {year} {2025})}\BibitemShut
  {NoStop}%
\bibitem [{\citenamefont {Wu}\ \emph {et~al.}(2023)\citenamefont {Wu},
  \citenamefont {Zhu}, \citenamefont {Bai}, \citenamefont {Wang},\ and\
  \citenamefont {Chiribella}}]{wu2023similarity}%
  \BibitemOpen
  \bibfield  {author} {\bibinfo {author} {\bibfnamefont {Y.-D.}\ \bibnamefont
  {Wu}}, \bibinfo {author} {\bibfnamefont {Y.}~\bibnamefont {Zhu}}, \bibinfo
  {author} {\bibfnamefont {G.}~\bibnamefont {Bai}}, \bibinfo {author}
  {\bibfnamefont {Y.}~\bibnamefont {Wang}},\ and\ \bibinfo {author}
  {\bibfnamefont {G.}~\bibnamefont {Chiribella}},\ }\href
  {https://doi.org/10.1103/PhysRevLett.130.210601} {\bibfield  {journal}
  {\bibinfo  {journal} {Physical Review Letters}\ }\textbf {\bibinfo {volume}
  {130}},\ \bibinfo {pages} {210601} (\bibinfo {year} {2023})}\BibitemShut
  {NoStop}%
\bibitem [{\citenamefont {Li}\ \emph {et~al.}(2026)\citenamefont {Li},
  \citenamefont {Zhan}, \citenamefont {Winston}, \citenamefont {Mer},
  \citenamefont {Yin}, \citenamefont {Yu}, \citenamefont {Alwehaibi},
  \citenamefont {Machado}, \citenamefont {Lopena}, \citenamefont {Zhang},
  \citenamefont {Kim}, \citenamefont {Zhang}, \citenamefont {Walmsley},\ and\
  \citenamefont {Patel}}]{li2026machine}%
  \BibitemOpen
  \bibfield  {author} {\bibinfo {author} {\bibfnamefont {Z.}~\bibnamefont
  {Li}}, \bibinfo {author} {\bibfnamefont {H.}~\bibnamefont {Zhan}}, \bibinfo
  {author} {\bibfnamefont {S.~H.}\ \bibnamefont {Winston}}, \bibinfo {author}
  {\bibfnamefont {E.}~\bibnamefont {Mer}}, \bibinfo {author} {\bibfnamefont
  {Z.}~\bibnamefont {Yin}}, \bibinfo {author} {\bibfnamefont {S.}~\bibnamefont
  {Yu}}, \bibinfo {author} {\bibfnamefont {Y.~K.}\ \bibnamefont {Alwehaibi}},
  \bibinfo {author} {\bibfnamefont {G.~J.}\ \bibnamefont {Machado}}, \bibinfo
  {author} {\bibfnamefont {D.~M.}\ \bibnamefont {Lopena}}, \bibinfo {author}
  {\bibfnamefont {L.}~\bibnamefont {Zhang}}, \bibinfo {author} {\bibfnamefont
  {M.~S.}\ \bibnamefont {Kim}}, \bibinfo {author} {\bibfnamefont
  {A.}~\bibnamefont {Zhang}}, \bibinfo {author} {\bibfnamefont {I.~A.}\
  \bibnamefont {Walmsley}},\ and\ \bibinfo {author} {\bibfnamefont {R.~B.}\
  \bibnamefont {Patel}},\ }\href {https://doi.org/10.48550/ARXIV.2602.23501}
  {\bibinfo {title} {Machine learning of quantum data using optimal similarity
  measurements}} (\bibinfo {year} {2026}),\ \Eprint
  {https://arxiv.org/abs/2602.23501} {arXiv:2602.23501} \BibitemShut {NoStop}%
\bibitem [{\citenamefont {Fawzi}\ \emph {et~al.}(2026)\citenamefont {Fawzi},
  \citenamefont {Oufkir},\ and\ \citenamefont {Salzmann}}]{fawzi2026optimal}%
  \BibitemOpen
  \bibfield  {author} {\bibinfo {author} {\bibfnamefont {O.}~\bibnamefont
  {Fawzi}}, \bibinfo {author} {\bibfnamefont {A.}~\bibnamefont {Oufkir}},\ and\
  \bibinfo {author} {\bibfnamefont {R.}~\bibnamefont {Salzmann}},\ }\href
  {https://doi.org/10.1103/qd1c-1fk9} {\bibfield  {journal} {\bibinfo
  {journal} {PRX Quantum}\ }\textbf {\bibinfo {volume} {7}},\ \bibinfo {pages}
  {010309} (\bibinfo {year} {2026})}\BibitemShut {NoStop}%
\bibitem [{\citenamefont {Gandhari}\ \emph {et~al.}(2024)\citenamefont
  {Gandhari}, \citenamefont {Albert}, \citenamefont {Gerrits}, \citenamefont
  {Taylor},\ and\ \citenamefont {Gullans}}]{gandhari2024precision}%
  \BibitemOpen
  \bibfield  {author} {\bibinfo {author} {\bibfnamefont {S.}~\bibnamefont
  {Gandhari}}, \bibinfo {author} {\bibfnamefont {V.~V.}\ \bibnamefont
  {Albert}}, \bibinfo {author} {\bibfnamefont {T.}~\bibnamefont {Gerrits}},
  \bibinfo {author} {\bibfnamefont {J.~M.}\ \bibnamefont {Taylor}},\ and\
  \bibinfo {author} {\bibfnamefont {M.~J.}\ \bibnamefont {Gullans}},\ }\href
  {https://doi.org/10.1103/prxquantum.5.010346} {\bibfield  {journal} {\bibinfo
   {journal} {PRX Quantum}\ }\textbf {\bibinfo {volume} {5}},\ \bibinfo {pages}
  {010346} (\bibinfo {year} {2024})}\BibitemShut {NoStop}%
\bibitem [{\citenamefont {Iosue}\ \emph {et~al.}(2024)\citenamefont {Iosue},
  \citenamefont {Sharma}, \citenamefont {Gullans},\ and\ \citenamefont
  {Albert}}]{iosue2024continuous}%
  \BibitemOpen
  \bibfield  {author} {\bibinfo {author} {\bibfnamefont {J.~T.}\ \bibnamefont
  {Iosue}}, \bibinfo {author} {\bibfnamefont {K.}~\bibnamefont {Sharma}},
  \bibinfo {author} {\bibfnamefont {M.~J.}\ \bibnamefont {Gullans}},\ and\
  \bibinfo {author} {\bibfnamefont {V.~V.}\ \bibnamefont {Albert}},\ }\href
  {https://doi.org/10.1103/physrevx.14.011013} {\bibfield  {journal} {\bibinfo
  {journal} {Physical Review X}\ }\textbf {\bibinfo {volume} {14}},\ \bibinfo
  {pages} {011013} (\bibinfo {year} {2024})}\BibitemShut {NoStop}%
\bibitem [{\citenamefont {Becker}\ \emph {et~al.}(2024)\citenamefont {Becker},
  \citenamefont {Datta}, \citenamefont {Lami},\ and\ \citenamefont
  {Rouze}}]{becker2024classical}%
  \BibitemOpen
  \bibfield  {author} {\bibinfo {author} {\bibfnamefont {S.}~\bibnamefont
  {Becker}}, \bibinfo {author} {\bibfnamefont {N.}~\bibnamefont {Datta}},
  \bibinfo {author} {\bibfnamefont {L.}~\bibnamefont {Lami}},\ and\ \bibinfo
  {author} {\bibfnamefont {C.}~\bibnamefont {Rouze}},\ }\href
  {https://doi.org/10.1109/tit.2024.3357972} {\bibfield  {journal} {\bibinfo
  {journal} {IEEE Transactions on Information Theory}\ }\textbf {\bibinfo
  {volume} {70}},\ \bibinfo {pages} {3427–3452} (\bibinfo {year}
  {2024})}\BibitemShut {NoStop}%
\bibitem [{\citenamefont {Thomas}\ \emph {et~al.}(2026)\citenamefont {Thomas},
  \citenamefont {Chabaud},\ and\ \citenamefont {Emeriau}}]{thomas2025shedding}%
  \BibitemOpen
  \bibfield  {author} {\bibinfo {author} {\bibfnamefont {H.}~\bibnamefont
  {Thomas}}, \bibinfo {author} {\bibfnamefont {U.}~\bibnamefont {Chabaud}},\
  and\ \bibinfo {author} {\bibfnamefont {P.-E.}\ \bibnamefont {Emeriau}},\
  }\href@noop {} {\bibfield  {journal} {\bibinfo  {journal}
  {\href{https://doi.org/10.1103/d6k4-4xrz}{PRX Quantum}}\ } (\bibinfo {year}
  {2026})},\ \bibinfo {note} {accepted for publication}\BibitemShut {NoStop}%
\bibitem [{\citenamefont {Mele}\ \emph {et~al.}(2025)\citenamefont {Mele},
  \citenamefont {Mele}, \citenamefont {Bittel}, \citenamefont {Eisert},
  \citenamefont {Giovannetti}, \citenamefont {Lami}, \citenamefont {Leone},\
  and\ \citenamefont {Oliviero}}]{mele2025}%
  \BibitemOpen
  \bibfield  {author} {\bibinfo {author} {\bibfnamefont {F.~A.}\ \bibnamefont
  {Mele}}, \bibinfo {author} {\bibfnamefont {A.~A.}\ \bibnamefont {Mele}},
  \bibinfo {author} {\bibfnamefont {L.}~\bibnamefont {Bittel}}, \bibinfo
  {author} {\bibfnamefont {J.}~\bibnamefont {Eisert}}, \bibinfo {author}
  {\bibfnamefont {V.}~\bibnamefont {Giovannetti}}, \bibinfo {author}
  {\bibfnamefont {L.}~\bibnamefont {Lami}}, \bibinfo {author} {\bibfnamefont
  {L.}~\bibnamefont {Leone}},\ and\ \bibinfo {author} {\bibfnamefont
  {S.~F.~E.}\ \bibnamefont {Oliviero}},\ }\href
  {https://doi.org/10.1038/s41567-025-03086-2} {\bibfield  {journal} {\bibinfo
  {journal} {Nature Physics}\ }\textbf {\bibinfo {volume} {21}},\ \bibinfo
  {pages} {2002} (\bibinfo {year} {2025})}\BibitemShut {NoStop}%
\bibitem [{\citenamefont {Daley}\ \emph {et~al.}(2012)\citenamefont {Daley},
  \citenamefont {Pichler}, \citenamefont {Schachenmayer},\ and\ \citenamefont
  {Zoller}}]{daley2012}%
  \BibitemOpen
  \bibfield  {author} {\bibinfo {author} {\bibfnamefont {A.~J.}\ \bibnamefont
  {Daley}}, \bibinfo {author} {\bibfnamefont {H.}~\bibnamefont {Pichler}},
  \bibinfo {author} {\bibfnamefont {J.}~\bibnamefont {Schachenmayer}},\ and\
  \bibinfo {author} {\bibfnamefont {P.}~\bibnamefont {Zoller}},\ }\href
  {https://doi.org/10.1103/PhysRevLett.109.020505} {\bibfield  {journal}
  {\bibinfo  {journal} {Physical Review Letters}\ }\textbf {\bibinfo {volume}
  {109}},\ \bibinfo {pages} {020505} (\bibinfo {year} {2012})}\BibitemShut
  {NoStop}%
\bibitem [{\citenamefont {Descamps}\ \emph {et~al.}(2026)\citenamefont
  {Descamps}, \citenamefont {Keller},\ and\ \citenamefont
  {Milman}}]{descamps2026}%
  \BibitemOpen
  \bibfield  {author} {\bibinfo {author} {\bibfnamefont {{\'E}.}~\bibnamefont
  {Descamps}}, \bibinfo {author} {\bibfnamefont {A.}~\bibnamefont {Keller}},\
  and\ \bibinfo {author} {\bibfnamefont {P.}~\bibnamefont {Milman}},\ }\href
  {https://doi.org/10.1103/jy6g-jp7n} {\bibfield  {journal} {\bibinfo
  {journal} {Physical Review Letters}\ }\textbf {\bibinfo {volume} {136}},\
  \bibinfo {pages} {060807} (\bibinfo {year} {2026})}\BibitemShut {NoStop}%
\bibitem [{\citenamefont {Liang}\ \emph {et~al.}(2019)\citenamefont {Liang},
  \citenamefont {Yeh}, \citenamefont {Mendon{\c{c}}a}, \citenamefont {Teh},
  \citenamefont {Reid},\ and\ \citenamefont {Drummond}}]{liang2019}%
  \BibitemOpen
  \bibfield  {author} {\bibinfo {author} {\bibfnamefont {Y.-C.}\ \bibnamefont
  {Liang}}, \bibinfo {author} {\bibfnamefont {Y.-H.}\ \bibnamefont {Yeh}},
  \bibinfo {author} {\bibfnamefont {P.~E. M.~F.}\ \bibnamefont
  {Mendon{\c{c}}a}}, \bibinfo {author} {\bibfnamefont {R.~Y.}\ \bibnamefont
  {Teh}}, \bibinfo {author} {\bibfnamefont {M.~D.}\ \bibnamefont {Reid}},\ and\
  \bibinfo {author} {\bibfnamefont {P.~D.}\ \bibnamefont {Drummond}},\ }\href
  {https://doi.org/10.1088/1361-6633/ab1ca4} {\bibfield  {journal} {\bibinfo
  {journal} {Reports on Progress in Physics}\ }\textbf {\bibinfo {volume}
  {82}},\ \bibinfo {pages} {076001} (\bibinfo {year} {2019})}\BibitemShut
  {NoStop}%
\bibitem [{\citenamefont {Weedbrook}\ \emph {et~al.}(2012)\citenamefont
  {Weedbrook}, \citenamefont {Pirandola}, \citenamefont
  {Garc{\'i}a-Patr{\'o}n}, \citenamefont {Cerf}, \citenamefont {Ralph},
  \citenamefont {Shapiro},\ and\ \citenamefont
  {Lloyd}}]{weedbrook2012gaussian}%
  \BibitemOpen
  \bibfield  {author} {\bibinfo {author} {\bibfnamefont {C.}~\bibnamefont
  {Weedbrook}}, \bibinfo {author} {\bibfnamefont {S.}~\bibnamefont
  {Pirandola}}, \bibinfo {author} {\bibfnamefont {R.}~\bibnamefont
  {Garc{\'i}a-Patr{\'o}n}}, \bibinfo {author} {\bibfnamefont {N.~J.}\
  \bibnamefont {Cerf}}, \bibinfo {author} {\bibfnamefont {T.~C.}\ \bibnamefont
  {Ralph}}, \bibinfo {author} {\bibfnamefont {J.~H.}\ \bibnamefont {Shapiro}},\
  and\ \bibinfo {author} {\bibfnamefont {S.}~\bibnamefont {Lloyd}},\ }\href
  {https://doi.org/10.1103/RevModPhys.84.621} {\bibfield  {journal} {\bibinfo
  {journal} {Reviews of Modern Physics}\ }\textbf {\bibinfo {volume} {84}},\
  \bibinfo {pages} {621} (\bibinfo {year} {2012})}\BibitemShut {NoStop}%
\bibitem [{\citenamefont {Robert}\ and\ \citenamefont
  {Casella}(2004)}]{robert2004montecarlointegration}%
  \BibitemOpen
  \bibfield  {author} {\bibinfo {author} {\bibfnamefont {C.~P.}\ \bibnamefont
  {Robert}}\ and\ \bibinfo {author} {\bibfnamefont {G.}~\bibnamefont
  {Casella}},\ }in\ \href {https://doi.org/10.1007/978-1-4757-4145-2_3} {\emph
  {\bibinfo {booktitle} {Monte Carlo Statistical Methods}}}\ (\bibinfo
  {publisher} {Springer},\ \bibinfo {address} {New York},\ \bibinfo {year}
  {2004})\ \bibinfo {edition} {2nd}\ ed.,\ pp.\ \bibinfo {pages}
  {79--122}\BibitemShut {NoStop}%
\bibitem [{\citenamefont {Klenke}(2020)}]{klenke2020probability}%
  \BibitemOpen
  \bibfield  {author} {\bibinfo {author} {\bibfnamefont {A.}~\bibnamefont
  {Klenke}},\ }\href {https://doi.org/10.1007/978-3-030-56402-5} {\emph
  {\bibinfo {title} {Probability Theory: A Comprehensive Course}}},\ \bibinfo
  {edition} {3rd}\ ed.\ (\bibinfo  {publisher} {Springer},\ \bibinfo {address}
  {Cham},\ \bibinfo {year} {2020})\BibitemShut {NoStop}%
\bibitem [{\citenamefont {Webb}\ \emph {et~al.}(2006)\citenamefont {Webb},
  \citenamefont {Ralph},\ and\ \citenamefont {Huntington}}]{webb2006homodyne}%
  \BibitemOpen
  \bibfield  {author} {\bibinfo {author} {\bibfnamefont {J.~G.}\ \bibnamefont
  {Webb}}, \bibinfo {author} {\bibfnamefont {T.~C.}\ \bibnamefont {Ralph}},\
  and\ \bibinfo {author} {\bibfnamefont {E.~H.}\ \bibnamefont {Huntington}},\
  }\href {https://doi.org/10.1103/PhysRevA.73.033808} {\bibfield  {journal}
  {\bibinfo  {journal} {Physical Review A}\ }\textbf {\bibinfo {volume} {73}},\
  \bibinfo {pages} {033808} (\bibinfo {year} {2006})}\BibitemShut {NoStop}%
\bibitem [{\citenamefont {Reck}\ \emph {et~al.}(1994)\citenamefont {Reck},
  \citenamefont {Zeilinger}, \citenamefont {Bernstein},\ and\ \citenamefont
  {Bertani}}]{reck1994experimental}%
  \BibitemOpen
  \bibfield  {author} {\bibinfo {author} {\bibfnamefont {M.}~\bibnamefont
  {Reck}}, \bibinfo {author} {\bibfnamefont {A.}~\bibnamefont {Zeilinger}},
  \bibinfo {author} {\bibfnamefont {H.~J.}\ \bibnamefont {Bernstein}},\ and\
  \bibinfo {author} {\bibfnamefont {P.}~\bibnamefont {Bertani}},\ }\href
  {https://doi.org/10.1103/PhysRevLett.73.58} {\bibfield  {journal} {\bibinfo
  {journal} {Physical Review Letters}\ }\textbf {\bibinfo {volume} {73}},\
  \bibinfo {pages} {58} (\bibinfo {year} {1994})}\BibitemShut {NoStop}%
\bibitem [{\citenamefont {Clements}\ \emph {et~al.}(2016)\citenamefont
  {Clements}, \citenamefont {Humphreys}, \citenamefont {Metcalf}, \citenamefont
  {Kolthammer},\ and\ \citenamefont {Walmsley}}]{clements2016optimal}%
  \BibitemOpen
  \bibfield  {author} {\bibinfo {author} {\bibfnamefont {W.~R.}\ \bibnamefont
  {Clements}}, \bibinfo {author} {\bibfnamefont {P.~C.}\ \bibnamefont
  {Humphreys}}, \bibinfo {author} {\bibfnamefont {B.~J.}\ \bibnamefont
  {Metcalf}}, \bibinfo {author} {\bibfnamefont {W.~S.}\ \bibnamefont
  {Kolthammer}},\ and\ \bibinfo {author} {\bibfnamefont {I.~A.}\ \bibnamefont
  {Walmsley}},\ }\href {https://doi.org/10.1364/OPTICA.3.001460} {\bibfield
  {journal} {\bibinfo  {journal} {Optica}\ }\textbf {\bibinfo {volume} {3}},\
  \bibinfo {pages} {1460} (\bibinfo {year} {2016})}\BibitemShut {NoStop}%
\bibitem [{\citenamefont {Knill}\ \emph {et~al.}(2001)\citenamefont {Knill},
  \citenamefont {Laflamme},\ and\ \citenamefont {Milburn}}]{knill2001scheme}%
  \BibitemOpen
  \bibfield  {author} {\bibinfo {author} {\bibfnamefont {E.}~\bibnamefont
  {Knill}}, \bibinfo {author} {\bibfnamefont {R.}~\bibnamefont {Laflamme}},\
  and\ \bibinfo {author} {\bibfnamefont {G.~J.}\ \bibnamefont {Milburn}},\
  }\href {https://doi.org/10.1038/35051009} {\bibfield  {journal} {\bibinfo
  {journal} {Nature}\ }\textbf {\bibinfo {volume} {409}},\ \bibinfo {pages}
  {46} (\bibinfo {year} {2001})}\BibitemShut {NoStop}%
\bibitem [{\citenamefont {Tang}\ \emph {et~al.}(2026)\citenamefont {Tang},
  \citenamefont {Zheng}, \citenamefont {Yu},\ and\ \citenamefont
  {Zhang}}]{tang2026distributed}%
  \BibitemOpen
  \bibfield {author} {\bibinfo {author} {\bibfnamefont {J.}~\bibnamefont
  {Tang}}, \bibinfo {author} {\bibfnamefont {C.}~\bibnamefont {Zheng}},
  \bibinfo {author} {\bibfnamefont {X.}~\bibnamefont {Yu}},\ and\
  \bibinfo {author} {\bibfnamefont {Z.}~\bibnamefont {Zhang}},\ }\href
  {https://arxiv.org/abs/2610.40141} {arXiv:2610.40141 [quant-ph] (2026)}
  \BibitemShut {NoStop}%
\end{thebibliography}
\end{document}